\pdfoutput=1  

\documentclass[aps,prd,superscriptaddress,floatfix,nofootinbib,preprintnumbers,eqsecnum,twocolumn,dvipsnames]{revtex4-1}

\usepackage{amsmath,float,graphicx,placeins,amssymb,multirow,pgfplots,pgfplotstable}
\usepackage{empheq}
\usepackage{tikz}
\usepackage{tikz-feynman}
\usepackage{subfigure}
\usepackage{comment}
\usepackage{enumerate,slashed,cancel,comment,color,bbm,graphicx,rotating,placeins,mathtools,multirow,hhline}
\usepackage[normalem]{ulem}
\usepackage{array}

\usepackage{hyperref}
\usepackage{color}
\usepackage{xcolor}
 \hypersetup
 {
   colorlinks,
   linkcolor={blue!80!black},
   citecolor={blue!70!black},
   urlcolor={blue!70!black}
 }

\usepackage{lipsum}
\usepackage{diagbox}
\usetikzlibrary{shapes.geometric,arrows}
\graphicspath{}
\usepackage{siunitx}
\usepackage{empheq}
\usepackage{xr}

\def\beq{\begin{equation}}
\def\eeq{\end{equation}}
\def\beqn{\begin{eqnarray}}
\def\eeqn{\end{eqnarray}}
\def\half{\mbox{\small ${\frac{1}{2}}$}}

\def\quarter{\mbox{\small ${\frac{1}{4}}$}}

\newcommand{\newc}{\newcommand}
\def\calZ{{\cal Z}}
\def\calM{{\cal M}}
\def\calV{{\cal V}}
\def\calF{{\cal F}}

\def\bQ{{\bf Q}}
\def\bT{{\bf T}}

\def\Qs{{\bf q}}

\def\KE{{\rm KE}}
\def\CM{{\rm CM}}

\def\barOmega{{\overline{\Omega}}}

\def\barXi{{\overline{\Xi}}}

\def\half{{\textstyle{1\over 2}}}
\def\quarter{{\textstyle{1\over 4}}}
\def\ie{{\it i.e.}\/}
\def\eg{{\it e.g.}\/}
\def\etc{{\it etc}.\/}
\def\inbar{\,\vrule height1.5ex width.4pt depth0pt}
\def\IR{\relax{\rm I\kern-.18em R}}
 \font\cmss=cmss10 \font\cmsss=cmss10 at 7pt
\def\IQ{\relax{\rm I\kern-.18em Q}}
\def\IZ{\relax\ifmmode\mathchoice
 {\hbox{\cmss Z\kern-.4em Z}}{\hbox{\cmss Z\kern-.4em Z}}
 {\lower.9pt\hbox{\cmsss Z\kern-.4em Z}}
 {\lower1.2pt\hbox{\cmsss Z\kern-.4em Z}}\else{\cmss Z\kern-.4em Z}\fi}

\renewcommand{\Im}{\textrm{Im}}

\newcommand {\eq}[1]{\mbox{Eq.~#1}}

\def\be{\begin{equation}}
\def\ee{\end{equation}}
\def\ba{\begin{eqnarray}}
\def\ea{\end{eqnarray}}

\def\W{\Omega}

\nonstopmode

\def\Tmin{T_{\rm min}}
\def\Tmax{T_{\rm max}}

\begin{document}

\title{
Cosmological Stasis from a Single Annihilating Particle Species: \\
Extending Stasis Into the Thermal Domain
}

\def\andname{\hspace*{-0.5em}} % gets rid of "and" in author list

\author{Jonah Barber}
%\email[Email address: ]{\tt jbarber2@arizona.edu}
\email[Email address: ]{jbarber2@arizona.edu}
\affiliation{Department of Physics, University of Arizona, Tucson, AZ 85721 USA}

\author{Keith R. Dienes}
%  \email[Email address: ]{\tt dienes@arizona.edu}
\email[Email address: ]{dienes@arizona.edu}
\affiliation{Department of Physics, University of Arizona, Tucson, AZ 85721 USA}
\affiliation{Department of Physics, University of Maryland, College Park, MD 20742 USA}

\author{Brooks Thomas}
%  \email[Email address: ]{\tt thomasbd@lafayette.edu}
\email[Email address: ]{thomasbd@lafayette.edu}
\affiliation{Department of Physics, Lafayette College, Easton, PA  18042 USA}

\begin{abstract}
It has recently been shown that extended cosmological epochs can exist during 
which the abundances associated with different energy components 
remain constant despite cosmological expansion.  Indeed, this ``stasis'' behavior 
has been found  to arise generically in many beyond-the-Standard-Model theories containing 
large towers of states, and even serves as a cosmological attractor.  However, all previous 
studies of stasis took place within non-thermal environments, or more specifically 
within environments in which thermal effects played no essential role in 
realizing or sustaining the stasis.  In this paper, we demonstrate that stasis 
can emerge and serve as an attractor even within thermal environments, with 
thermal effects playing a critical role in the stasis dynamics.    
Moreover, within such environments, we find that no towers of states are needed ---  
a single state experiencing two-body annihilations will suffice.  This work thus 
extends the stasis phenomenon into the thermal domain and demonstrates that thermal 
effects can also generally give rise to an extended stasis epoch, even 
when only a single matter species is involved.
\end{abstract}
\maketitle

\tableofcontents

\def\ie{{\it i.e.}\/}
\def\eg{{\it e.g.}\/}
\def\etc{{\it etc}.\/}
\def\taubar{{\overline{\tau}}}
\def\qbar{{\overline{q}}}
\def\kbar{{\overline{k}}}
\def\bQ{{\bf Q}}
\def\calT{{\cal T}}
\def\calN{{\cal N}}
\def\calF{{\cal F}}
\def\calM{{\cal M}}
\def\calZ{{\cal Z}}

\def\beq{\begin{equation}}
\def\eeq{\end{equation}}
\def\beqn{\begin{eqnarray}}
\def\eeqn{\end{eqnarray}}
\def\apo{\mbox{\small ${\frac{\alpha'}{2}}$}}
\def\half{\mbox{\small ${\frac{1}{2}}$}}
\def\sqapo{\mbox{\tiny $\sqrt{\frac{\alpha'}{2}}$}}
\def\sqap{\mbox{\tiny $\sqrt{{\alpha'}}$}}
\def\sqapxtwo{\mbox{\tiny $\sqrt{2{\alpha'}}$}}
\def\aptwo{\mbox{\tiny ${\frac{\alpha'}{2}}$}}
\def\apofour{\mbox{\tiny ${\frac{\alpha'}{4}}$}}
\def\bosqtwo{\mbox{\tiny ${\frac{\beta}{\sqrt{2}}}$}}
\def\btosqtwo{\mbox{\tiny ${\frac{\tilde{\beta}}{\sqrt{2}}}$}}
\def\apofour{\mbox{\tiny ${\frac{\alpha'}{4}}$}}
\def\sqaptwo{\mbox{\tiny $\sqrt{\frac{\alpha'}{2}}$}  }
\def\apoeight{\mbox{\tiny ${\frac{\alpha'}{8}}$}}
\def\sapoeight{\mbox{\tiny ${\frac{\sqrt{\alpha'}}{8}}$}}

\newc{\gsim}{\lower.7ex\hbox{{\mbox{$\;\stackrel{\textstyle>}{\sim}\;$}}}}
\newc{\lsim}{\lower.7ex\hbox{{\mbox{$\;\stackrel{\textstyle<}{\sim}\;$}}}}
\def\calM{{\cal M}}
\def\calV{{\cal V}}
\def\calF{{\cal F}}
\def\bQ{{\bf Q}}
\def\bT{{\bf T}}
\def\Qs{{\bf q}}

\def\half{{\textstyle{1\over 2}}}
\def\quarter{{\textstyle{1\over 4}}}
\def\ie{{\it i.e.}\/}
\def\eg{{\it e.g.}\/}
\def\etc{{\it etc}.\/}
\def\inbar{\,\vrule height1.5ex width.4pt depth0pt}
\def\IR{\relax{\rm I\kern-.18em R}}
 \font\cmss=cmss10 \font\cmsss=cmss10 at 7pt
\def\IQ{\relax{\rm I\kern-.18em Q}}
\def\IZ{\relax\ifmmode\mathchoice
 {\hbox{\cmss Z\kern-.4em Z}}{\hbox{\cmss Z\kern-.4em Z}}
 {\lower.9pt\hbox{\cmsss Z\kern-.4em Z}}
 {\lower1.2pt\hbox{\cmsss Z\kern-.4em Z}}\else{\cmss Z\kern-.4em Z}\fi}

%=====================================================================================================

%%%%%%%%%%%%%%%%%%%%%%%%%%%%%%%%%%%%%%%%%%%%%%%%%%%%%%%%%%%%%%%%%%%%%%%%%%%%%%

\section{Introduction\label{sec:intro}}

%%%%%%%%%%%%%%%%%%%%%%%%%%%%%%%%%%%%%%%%%%%%%%%%%%%%%%%%%%%%%%%%%%%%%%%%%%%%%%

It has recently been observed that many theories of physics beyond the 
Standard Model (BSM) give rise to early-universe cosmologies which potentially 
contain extended epochs during which the abundances of different cosmological 
energy components (such as matter, radiation, and vacuum energy) 
remain constant despite cosmological expansion.  This phenomenon has been dubbed
``stasis''~\cite{Dienes:2021woi}, and seems to be rather ubiquitous in systems 
containing large (or even infinite) towers of increasingly massive states, such 
as are known to arise in many BSM
models~\cite{Barrow:1991dn,Dienes:2021woi,Dienes:2022zgd,Dienes:2023ziv,
Dienes:2024wnu,Halverson:2024oir}.  
Indeed, it has been found that matter and radiation can be in stasis with each
other~\cite{Dienes:2021woi,Barrow:1991dn, Dienes:2022zgd,Halverson:2024oir}, and 
that each of these can also be in stasis with vacuum 
energy~\cite{Dienes:2023ziv,Dienes:2024wnu}.   It has even been found that 
matter, radiation, and vacuum energy can all experience a simultaneous 
{\it triple stasis}\/~\cite{Dienes:2023ziv}. Moreover, 
stasis is a dynamical attractor within such cosmologies.  Thus, even if these systems 
do not begin in stasis, they will naturally flow towards such stasis configurations.   
This renders the stasis state essentially unavoidable in many BSM cosmologies.

All of the BSM cosmologies which have been examined thus far share certain 
characteristics.   First, they are all non-thermal.   By this, we mean that they 
do not involve any cosmological energy components whose temperatures play a critical 
role in the stasis phenomenon.  Even in cases for which the stasis is realized through the 
Hawking radiation emitted by primordial black holes~\cite{Barrow:1991dn,Dienes:2022zgd}, 
it is sufficient to regard the black-hole evaporation process as one which simply 
converts matter (in the form of black holes) to radiation.  
In such analyses --- indeed, in all of the different stasis scenarios examined 
thus far --- the temperatures that might happen to characterize particular cosmological 
populations of particles play no essential role in the stasis dynamics.   
Secondly, and perhaps even more importantly, in all of the BSM systems examined thus far   
the existence of a {\it bona fide}\/ tower 
of different species $\phi_\ell$ with different masses $m_\ell$ was critical.   
Indeed, we have even repeatedly found that the duration of the resulting stasis 
epoch is directly related to the number of species in the tower.  Finally --- and 
a bit more technically --- the energy-transfer mechanisms that drove the stasis 
phenomenon in each of the previous cases (such as the decay of matter to 
radiation in Refs.~\cite{Dienes:2021woi, Barrow:1991dn, Dienes:2023ziv}, 
or the transition from overdamped fields to underdamped fields in 
Refs.~\cite{Dienes:2023ziv,Dienes:2024wnu}) had the property that the {\it rate}\/ of
energy transfer depended linearly on the energy density associated with the original 
field.  More specifically, the ``pumps'' $P^{(\rho)}$ that were required in each of 
the previous cases were each proportional to  the corresponding energy density $\rho$.  
This linearity property played 
a significant role in establishing the algebraic structure of the resulting stasis.  
While suitable for decay processes in which a matter particle $\phi$ decays into 
radiation, thereby leading to a rate of energy transfer from matter to radiation scaling
linearly with the $\phi$-particle density $n_\phi$, this linearity property would seem 
to preclude processes such as particle annihilation, for which we would expect a rate that 
scales as $n_\phi^2$.

Given these observations, one might suspect that each of these features plays a 
necessary role in establishing the stasis phenomenon.  However, in this paper, we shall 
demonstrate that {\it none of these features are actually required for stasis}\/.
In particular, we shall describe an explicit thermal scenario
which has none of the features outlined 
above, but which nevertheless is capable of producing an epoch of stasis persisting 
across many {\it e}\/-folds of cosmological expansion.  
Indeed, we shall even show that such a stasis epoch is a cosmological attractor.

As might be imagined, the thermal stasis that we shall describe in this paper is 
quite different from the non-thermal stases that already exist in the literature.   
Without a tower of states, and with the stasis driven by particle annihilation 
rather than by particles decaying or undergoing phase transitions, it might seem 
at first glance that thermal stasis must emerge in a completely different way.   
Indeed, even the pumps that we shall employ in this paper may be more familiar from studies of 
thermal freeze-out in weakly-interacting-massive-particle (WIMP) models of dark matter
than from prior studies of stasis.  However, despite these important differences, 
we shall nevertheless find that a stasis likewise emerges within this new context and shares 
many critical properties with its non-thermal cousins.

This paper is organized as follows.  In Sect.~\ref{sec:preliminaries}, we describe the 
different cosmological energy components that will be present in our thermal-stasis framework.
We also discuss the fundamental assumptions regarding the manner in which they behave.  
As we shall see, one of these components is a population of non-relativistic particles 
which are in kinetic equilibrium with each other.  Then, in Sect.~\ref{sec:dynamics}, 
we derive the evolution equations not only for the cosmological abundances of our energy 
components but also for the temperature of this non-relativistic particle gas.  In
Sect.~\ref{sec:pumps}, we examine the general forms of the pump terms that arise in these 
equations due to the annihilation of our non-relativistic particles. In 
Sect.~\ref{sec:coldness}, we then proceed to identify a particular combination of our 
dynamical variables --- a combination which we call ``coldness''  --- which plays a 
crucial role in the cosmological dynamics, and we recast our evolution equations in terms 
of this new variable.  In Sect.~\ref{sec:fixed_point} we examine the fixed-point solutions
to these equations, and demonstrate that a novel form of stasis emerges as such a fixed point
under certain conditions.  Indeed, we shall find that it is the coldness rather than the 
temperature which remains constant during thermal stasis. In Sect.~\ref{sec:attractor}, 
we demonstrate that this stasis solution is not merely a local attractor, but actually 
a {\it global}\/ attractor when these conditions are satisfied.
In Sect.~\ref{sec:mechanism_q_neg_2}, we then identify a particular mechanism through which 
these stasis conditions can naturally be realized in a particle-physics context. 
Finally, in Sect.~\ref{sec:conclusions}, we conclude with a summary of our results and a 
discussion highlighting possible directions for future work.

%%%%%%%%%%%%%%%%%%%%%%%%%%%%%%%%%%%%%%%%%%%%%%%%%%%%%%%%%%%%%%%%%%%%%%%%%%%%%%

\section{Assumptions and Energy Components\label{sec:preliminaries}}

%%%%%%%%%%%%%%%%%%%%%%%%%%%%%%%%%%%%%%%%%%%%%%%%%%%%%%%%%%%%%%%%%%%%%%%%%%%%%%

In this paper, our goal is to study the extent to which stasis can emerge in a thermal 
environment.  Toward this end, in this section we shall begin by describing the cosmological 
energy components present in our framework for establishing such a stasis 
and the fundamental assumptions regarding the manner in which they behave.  
For sake of generality, we shall keep the discussion as model-independent as possible in
what follows.  Later, in Sect.~\ref{sec:mechanism_q_neg_2}, we shall present an explicit 
mechanism through which these conditions can be realized in a particle-physics context.

Let us start by considering the cosmology associated with a universe which comprises 
a non-relativistic matter species $\phi$ of mass $m$ which annihilates into an effectively 
massless particle $\chi$ through processes of the form $\phi\phi\to \chi\chi$.  For simplicity, 
we shall assume that $\phi$ and $\chi$ are both their own antiparticles (although our qualitative 
results would be unchanged if we were to drop this last assumption).
Thus, such a universe contains not only a matter component $\phi$ with an energy density $\rho_M$, 
abundance $\Omega_M$, and equation-of-state parameter $w_M=0$, but also a radiation 
component $\chi$ with a corresponding energy density $\rho_\gamma$, abundance $\Omega_\gamma$, 
and equation-of-state parameter $w_\gamma=1/3$.  For simplicity, we shall assume that $\phi$ 
and $\chi$ are the only cosmological species  present 
within our universe, and that the universe is flat and expanding in a 
manner consistent with the Friedmann-Robertson-Walker (FRW) metric.

Remarkably, we shall demonstrate that such a universe can give rise to an extended 
stasis epoch.   To do this,  we shall assume that our population of matter particles 
comprises only a single species $\phi$ with a unique mass $m$, as noted above.  We shall also assume that this 
species is stable in the sense that it does not decay within the time frame relevant for our 
analysis.  Finally, we shall assume that the $\phi$ particles are not absolutely cold, but 
rather are in {\it thermal equilibrium}\/ with each other, possibly as the result of rapid 
interactions among them, and constitute an ideal gas with a non-zero temperature $T$.   
 
Note that our assumption that the $\phi$ particles form an ideal gas does {\it not}\/ imply 
the existence of a thermal bath which holds the temperature of this gas constant.
Rather, we shall allow the temperature of our $\phi$-particle gas to evolve dynamically along with 
the rest of the cosmology.  In other words, while we will assume the existence of rapid 
interactions {\it amongst}\/ the $\phi$ particles which ensure that they remain in thermal 
equilibrium {\it with each other}\/, we are not assuming that these particles have additional 
interactions with any other particle species that might have constituted a thermal bath.

Given these assumptions, we see that we must actually regard our universe as containing 
{\it three}\/ different ``fluids.''  The first two fluids are those mentioned above, namely 
those associated with matter and radiation, but our matter fluid really only accounts for the 
{\it rest-mass energy} of our matter fields.   We must therefore now introduce a third fluid, namely 
that associated with the {\it kinetic energy}\/ of our matter fields, with corresponding energy 
density $\rho_\KE$, abundance $\Omega_\KE$, and equation-of-state parameter $w_\KE$.  It is 
important to note that we cannot consider $\rho_M$ and $\rho_\KE$ as different contributors to a 
single common fluid associated with our gas of $\phi$ particles with a unique time-independent 
equation of state of the form $P=w\rho$ with a constant $w$, since (as we shall demonstrate
below) the energy densities associated with rest mass and kinetic energy have different 
equations of state of this form, with $w_M\not= w_\KE$.
 
Since we are assuming that our $\phi$ particles constitute a non-relativistic ideal gas at 
temperature $T$, it follows that $\rho_\KE$ and $T$ are related to each other via 
\beq 
  \rho_\KE ~=~ \frac{3}{2} n_M T ~\approx~ \frac{3}{2} \rho_M \frac{T}{m}~,
  \label{idealgas}
\eeq
where $n_M \approx \rho_M/m$ is the number density of $\phi$ particles in the gas.
The kinetic-energy abundance is therefore
\beq
  \Omega_\KE ~=~ \frac{3}{2} \Omega_M \frac{T}{m}~.
  \label{KETrelation}
\eeq
Thus, for a given $\Omega_M$, we can trade $\Omega_\KE$ for $T$ and {\it vice versa}\/.

We can also define an equation-of-state parameter $w_\KE$ for the kinetic energy of our 
$\phi$-particle gas.  The ideal gas law tells us that $P=n_M T$, where $P$ is the total 
pressure associated with this gas.  It then follows from Eq.~(\ref{idealgas}) that 
$\rho_\KE = 3P/2$, which implies that
\beq
    w_\KE ~=~ \frac{2}{3}~.
\label{w_for_KE}
\eeq
Another way to understand this result is to consider how the kinetic-energy density of our 
$\phi$ particles scales with the scale factor $a$.  In general, the kinetic-energy density
of a localized population of particles in a FRW universe always includes a factor $a^{-3}$ 
due to the expansion of the volume within which those particles are contained.  However, 
this kinetic-energy density may also accrue an additional dependence on $a$ from the manner 
in which the kinetic energies of the individual particles are affected by the redshifting of 
their momenta.  For example, if the particle is highly relativistic, its kinetic energy
is approximately proportional to its momentum $p$ and therefore scales as $a^{-1}$.  
Thus, the kinetic-energy density of a population of such a particles --- which is 
essentially equivalent to their total energy density --- scales as $a^{-4}$.  
Since we generally have $\rho\sim a^{-3(1+w)}$ for a perfect fluid with a constant 
equation-of-state parameter $w$, we obtain the usual result $w = 1/3$ for radiation.
By contrast, the kinetic energy associated with a {\it non}\/-relativistic particle
is proportional to $p^2$ and therefore scales as $a^{-2}$.  Thus, the kinetic-energy
density of a population of non-relativistic particles scales as $a^{-5}$.  This 
corresponds to an equation-of-state parameter $w= 2/3$, in accordance with 
Eq.~(\ref{w_for_KE}).

%%%%%%%%%%%%%%%%%%%%%%%%%%%%%%%%%%%%%%%%%%%%%%%%%%%%%%%%%%%%%%%%%%%%%%%%%%%%%%

\section{Dynamical Equations\label{sec:dynamics}}

%%%%%%%%%%%%%%%%%%%%%%%%%%%%%%%%%%%%%%%%%%%%%%%%%%%%%%%%%%%%%%%%%%%%%%%%%%%%%%

It is now relatively straightforward to derive the equations that govern the time-evolution 
of the three abundances $\Omega_M$, $\Omega_\gamma$, and $\Omega_\KE$ within this cosmology.
Each of these abundances $\Omega_i$ is related to the corresponding energy density 
$\rho_i$ via
\beq
  \Omega_i   ~\equiv~ \frac{8\pi G}{3H^2} \,\rho_i~~
\label{eq:Omegadef}
\eeq
where $i\in \lbrace M,\gamma,\KE\rbrace$, where $H\equiv \dot a/a$ is the Hubble parameter, 
and where $G$ is Newton's constant.
It then follows that
\beq
  \frac{d\Omega_i}{dt} ~=~ \frac{8\pi G}{3} \left( \frac{1}{H^2} \frac{d\rho_i}{dt} - 2 
    \frac{\rho_i}{H^3} \frac{dH}{dt} \right)~.
\label{stepone}
\eeq
On the other hand, for our three-component universe, the Friedmann acceleration 
equation tells us that
\beqn
  \frac{dH}{dt}  
  ~&=&~ -H^2 - \frac{4\pi G}{3} \left(  \sum_i \rho_i + 3 \sum_i p_i\right)\nonumber\\
  ~&=&~ - \half H^2 \left( 2+ \Omega_M + 2 \Omega_\gamma + 3 \Omega_\KE\right)\nonumber\\
  ~&=&~ - \half H^2 \left( 4-\Omega_M + \Omega_\KE\right)~
\label{acceleq}
\eeqn
where in passing from the second to the third line we have imposed the constraint 
$\Omega_M + \Omega_\gamma + \Omega_\KE=1$, as befits our three-component universe.
Substituting Eq.~(\ref{acceleq}) into Eq.~(\ref{stepone}) then yields
\beqn
    \frac{d\Omega_M}{dt} 
     ~&=&~ \frac{8\pi G}{3H^2} \,\frac{d\rho_M}{dt} 
     + H \Omega_M \left( 4-\Omega_M + \Omega_\KE \right) ~ \nonumber\\
    \frac{d\Omega_\gamma}{dt} 
     ~&=&~ \frac{8\pi G}{3H^2} \,\frac{d\rho_\gamma}{dt}  
    + H \Omega_\gamma \left( 4-\Omega_M + \Omega_\KE\right)~\nonumber\\
    \frac{d\Omega_\KE}{dt} 
     ~&=&~ \frac{8\pi G}{3H^2} \,\frac{d\rho_\KE}{dt}  
    + H \Omega_\KE \left( 4-\Omega_M + \Omega_\KE\right)~.~\nonumber\\
\label{convert2}
\eeqn

In general, the rates of change $d\rho_M/dt$, $d\rho_\gamma/dt$, and $d\rho_\KE/dt$ 
of the energy densities associated with our three cosmological components may 
be written schematically as
\beqn
  \frac{d\rho_M}{dt} ~&=&~ -3 H \rho_M - P^{(\rho)}_{M,\gamma} ~ \nonumber\\
  \frac{d\rho_\gamma}{dt} ~&=&~ -4 H \rho_\gamma + P^{(\rho)}_{M,\gamma} +
    P^{(\rho)}_{\KE,\gamma}~ \nonumber\\
    \frac{d\rho_\KE}{dt} ~&=&~ -5 H \rho_\KE - P^{(\rho)}_{\KE,\gamma}~, 
\label{eq:eoms}
\eeqn
where the first term on the right side of each equation represents the 
effect of cosmological expansion and where each of the remaining terms 
$P^{(\rho)}_{ij}$ represents the rate at which energy-density is transferred 
from component $i$ to component $j$ as a consequence of our annihilation 
process $\phi\phi\to\chi\chi$.  In general, this annihilation process eliminates $\phi$ 
particles and thereby reduces not only the rest-mass energy density $\rho_M$ of the 
$\phi$-particle gas but also its kinetic energy density $\rho_\KE$.  The fact 
that this process conserves energy then implies that any annihilation-induced 
reductions in $\rho_M$ and $\rho_\KE$ must lead to a corresponding increase in 
$\rho_\gamma$.  

Note that we have not included similar terms for the inverse process 
$\chi\chi \to \phi\phi$ in \eq(\ref{eq:eoms}).  However, the energy-density-transfer 
rates associated with this process are generally negligible within our primary regime 
of interest.  To see this, we begin by noting that since our gas of 
$\phi$ particles is presumed to be highly non-relativistic, the energies $E_\chi$ 
of the $\chi$ particles produced by $\phi\phi\to\chi\chi$ annihilation at any given time 
are initially sharply peaked around $E_\chi \sim m$ in the cosmological background frame, with 
a width $\Delta E_\chi \sim \sqrt{mT} \ll m$ where $T$ is the temperature of the $\phi$-particle 
gas at that time.  However, these energies subsequently decrease as a result of cosmological 
redshifting.  In situations in which the $\chi$ particles interact sufficiently weakly that 
little redistribution of the $E_\chi$ takes place after they are produced, these energies 
quickly fall below the kinematic threshold for $\chi\chi\to \phi\phi$ production.  Even in 
situations in which the $\chi$ particles are more strongly interacting and a more significant 
redistribution of $E_\chi$ values takes place, only a small (and ever-decreasing) 
fraction of $\chi$-particle pairs will ultimately have energies above this threshold.  
In either case, then, cosmological redshifting effectively renders the swept-volume rate 
for $\chi\chi\to \phi\phi$ production negligible.

We also note that while the scattering process $\chi\phi \to \chi\phi$ can in principle 
affect the distribution of kinetic energies for the $\phi$-particle gas, the swept-volume 
rate for this process need not be comparable to the swept-volume rate for 
$\phi\phi\to \chi\chi$ annihilation.  For example, if 
$\phi\phi\to\chi\chi$ annihilation were to proceed through an $s$-channel process which 
occurs on resonance while simultaneously $\chi\phi\to \chi\phi$ were to occur only through 
$t$-channel processes, the swept-volume rate for the former process could be significant
while that for the latter process could be sufficiently small that its impact on the 
dynamical evolution of the system could be neglected.  Indeed, as we shall see in
Sect.~\ref{sec:mechanism_q_neg_2}, there exist scenarios for thermal stasis in which this 
is in fact the case.  We shall therefore assume in what follows that the effect of 
$\chi\phi\to \chi\phi$ scattering on the kinetic-energy distribution of the $\phi$ particles
can be neglected.

In general, following Ref.~\cite{Dienes:2023ziv}, we shall refer to any process 
that induces a transfer of energy between two energy components as a ``pump.''  
Such pumps are associated with local particle-physics processes which conserve 
energy density, but redistribute it among our different energy components.

Substituting the expressions in \eq(\ref{eq:eoms}) into Eq.~(\ref{convert2}), we have
\beqn
  \frac{d\Omega_M}{dt} ~&=&~ 
    H \Omega_M\left( 1 - \Omega_M + \Omega_\KE\right)~ -P_{M,\gamma} \nonumber\\
  \frac{d\Omega_\gamma}{dt} ~&=&~ 
    H \Omega_\gamma \left(- \Omega_M + \Omega_\KE\right) 
    + P_{M,\gamma}+P_{\KE,\gamma} ~\nonumber\\
  \frac{d\Omega_\KE}{dt} ~&=&~ 
    H \Omega_\KE\left( -1 - \Omega_M + \Omega_\KE\right)~ -P_{\KE,\gamma}~,~~~~~
\label{eq:convert3}
\eeqn
where we have defined
\beq
  P_{ij} ~\equiv~ \frac{8\pi G}{3 H^2} \, P^{(\rho)}_{ij}~.
  \label{eq:abundance_pump}
\eeq
Indeed, while the pump term $P^{(\rho)}_{ij}$ represents a transfer of {\it energy density}\/, 
the corresponding pump term $P_{ij}$ represents a transfer of {\it abundance}\/.
We observe from Eq.~(\ref{eq:convert3}) that $d\Omega_M/dt+ d\Omega_\gamma/dt+ d\Omega_\KE/dt=0$, 
which reflects the fact that $\Omega_M+\Omega_\gamma+\Omega_\KE=1$ within this three-component 
universe.  For this reason, we shall henceforth refrain from writing down explicit expressions 
for $d\Omega_\gamma/dt$, as these expressions can always be determined 
directly from $d\Omega_M/dt$ and $d\Omega_\KE/dt$.

From our result for $d\Omega_\KE/dt$ in \eq(\ref{eq:convert3}) and the relation in 
Eq.~(\ref{KETrelation}), we may also obtain an expression for the rate of change of the 
temperature of the $\phi$ particles.  This expression takes the form
\beq
  \frac{dT}{dt} ~=~
    -2 H T  - 
    \frac{2m}{3\Omega_M} 
    \left( P_{\KE,\gamma} 
    - \frac{\Omega_\KE}{\Omega_M}
    \,P_{M,\gamma} \right)~.
  \label{eq:dTdtequation}
\eeq
This result implies that while the pump terms $P_{\KE,\gamma}$ and $P_{M,\gamma}$ both serve 
to decrease the overall energy density associated with our $\phi$-particle gas, 
the former term serves to decrease its temperature while the latter serves to increase it.
Indeed, we observe that the net effect of the pumps is to raise $T$ when  
\beq 
  \frac{P_{\KE,\gamma}}{\Omega_\KE} ~<~
  \frac{P_{M,\gamma}}{\Omega_M} ~,
  \label{pumpbalance}
\eeq
and to lower $T$ if the opposite is true.  Interestingly, if both sides of 
Eq.~(\ref{pumpbalance}) are equal, these pumps will have no net effect on $T$.  
Of course, $T$ still decreases as a consequence of cosmological expansion in this case, 
as indicated by the first term on the right side of Eq.~(\ref{eq:dTdtequation}).

%%%%%%%%%%%%%%%%%%%%%%%%%%%%%%%%%%%%%%%%%%%%%%%%%%%%%%%%%%%%%%%%%%%%%%%%%%%%%%

\section{Pump Terms for Thermal Annihilation\label{sec:pumps}}

%%%%%%%%%%%%%%%%%%%%%%%%%%%%%%%%%%%%%%%%%%%%%%%%%%%%%%%%%%%%%%%%%%%%%%%%%%%%%%

It is not difficult to obtain 
explicit expressions for our pump terms.  Within the cosmology we are studying, such 
pumps represent the effects of the two-body annihilations of $\phi$ particles into radiation.   
Such process are familiar from traditional studies of the thermal freeze-out phenomenon, 
where one has $dn_M/dt\sim \langle \sigma v\rangle n_M^2 +...$ where $n_M$ is the 
matter-particle number density.  In this expression, $\langle \sigma v\rangle$ denotes 
the thermally averaged ``swept-volume'' rate, \ie, the rate at which volume is swept 
by the cross-section $\sigma$ moving with transverse velocity $v$. 
Recognizing that $n_M=\rho_M/m$, we see that our corresponding pump is nothing but
\begin{equation}
    P^{(\rho)}_{M,\gamma} ~=~ \frac{1}{m}\langle \sigma v\rangle \rho_M^2~.
\end{equation}
We stress that 
in writing this relation we have implicitly assumed that our population of $\phi$ particles 
is in thermal equilibrium with itself.  Likewise, the assumption that our matter is 
non-relativistic implies that $ T\ll m$ where $T$ is the temperature of our ideal 
gas of $\phi$ particles. 

The same annihilation process also acts to decrease the kinetic-energy density of
$\phi$-particle gas.
We find that the corresponding kinetic-energy pump is given in terms of the 
thermally averaged kinetic-energy-weighted cross-section by 
\begin{equation}
  P^{(\rho)}_{\KE,\gamma} ~=~ 
    \frac{1}{2m^2}\Bigl\langle \bigl(\KE_a + \KE_b\bigr)
    \sigma v\Bigr\rangle \rho_M^2~
\end{equation}
where $\KE_a$ and $\KE_b$ are the kinetic energies of the annihilating $\phi$ particles.
Inserting the results for the corresponding pumps $P_{M,\gamma}$ and $P_{\KE,\gamma}$ 
into Eq.~(\ref{eq:dTdtequation}), we have
\beq
  \frac{dT}{dt} = -2 H T + \frac{\rho_M T}{m} 
    \,\langle \sigma v\rangle  - \frac{\rho_M}{3m} 
    \Bigl\langle \left(\KE_a + \KE_b\right)\sigma v \Bigr\rangle~.
\label{eq:TempDiffEq}
\eeq

In order to evaluate these thermally averaged cross-sections
$\langle \sigma v\rangle$ and $\langle (\KE_a+\KE_b) \sigma v\rangle$, we need to know 
how $\sigma v$ depends on the incoming particle momentum  
as seen within the center-of-mass (CM) frame.  Indeed, since our population 
of $\phi$ particles is assumed to be highly non-relativistic, the CM frame
and the cosmological background frame approximately coincide.
We shall make an {\it ansatz}\/ and assume that $\sigma v$ can be 
parametrized as 
\beq
  \sigma v ~=~ C\, \left(\frac{|\vec p_\CM|}{m}\right)^q~,
  \label{eq:swept_vol_rate}
\eeq
where $|\vec p_\CM | \equiv |\vec p_a-\vec p_b|$ is the magnitude of the 
three-momentum of either of the two incoming particles $\phi_a$ and $\phi_b$ 
in the CM frame, where $q$ is an arbitrary exponent, and where $C$ is a 
momentum-independent and $q$-independent coefficient.
The ansatz in Eq.~(\ref{eq:swept_vol_rate}) is not entirely unfamiliar;
for example, a similar ansatz is often invoked when discussing 
thermal freeze-out (see, \eg, Ref.~\cite{Kolb:1990vq,Jungman:1995df,Bertone:2004pz}), 
with the exponent $q$ selecting between $s$-wave annihilation 
($q=0$), $p$-wave annihilation ($q=2$), $d$-wave annihilation ($q=4$), and so forth.   
However, although the value of $q$ na\"\i vely corresponds to twice the order of the 
leading term in the partial-wave expansion, $q$ can actually be more negative than 
stated above if our annihilation proceeds through an $s$-channel process with a propagator 
that is in resonance at low momenta.  Furthermore, Sommerfeld 
enhancement~\cite{Sommerfeld:1931qaf} can also give rise to negative values of $q$.
For this reason, we will also allow ourselves to 
consider values of $q$ which are negative.  Indeed, we shall give an example of a 
model that satisfies Eq.~(\ref{eq:swept_vol_rate}) with such values of $q$ in 
Sect.~\ref{sec:mechanism_q_neg_2}. 

Given the ansatz in Eq.~(\ref{eq:swept_vol_rate}), we can now evaluate our thermally 
averaged cross-sections in terms of $C$ and $q$.  By definition, these cross-sections 
can be written as
\beq
\langle \sigma v\rangle \, =\, 
   \frac{C}{m^q} \int d^3 \vec p_a \int d^3 \vec p_b ~ \left(
   \frac{|\vec p_a-\vec p_b|}{2} \right)^q f(\vec p_a)f(\vec p_b)~\\ 
\label{eq:cross1}
\eeq
and 
\beqn
&& \left\langle \bigl (\KE_a+\KE_b\bigr )\sigma v\right\rangle  \,=\,
   \frac{C}{m^q} \int d^3 \vec p_a\! \int d^3 \vec p_b \, 
      \left(
   \frac{|\vec p_a-\vec p_a|}{2} \right)^q~~\,\nonumber\\
&&  ~~~~~~~~~~~~~ \times\,
   \left( \frac{|\vec p_a|^2+|\vec p_b| ^2}{2m}\right)
   \, f(\vec p_a)f(\vec p_b)
\label{eq:cross2}
\eeqn
where the normalized thermal (Boltzmann) suppression factor for 
each of our two annihilating $\phi$ particles is given by
\beq
   f(\vec{p}\,) ~\equiv~ \frac{1}{(2\pi m T)^{3/2}}\,
      \exp\left ( - \frac{|\vec{p}\,|^2}{2mT}\right)~.
\label{eq:thermalf}
\eeq
We immediately see that we must have $q> -3$ in order for the integrals 
appearing in Eqs.~(\ref{eq:cross1}) and (\ref{eq:cross2}) to be free of infrared 
divergences.  For values of $q$ which satisfy this criterion, we may evaluate these 
integrals in a straightforward manner by performing a change of variables from 
$\vec p_a$ and $\vec p_b$ to $\vec p_+$ and $\vec p_-$, where 
$\vec p_\pm\equiv \vec p_a\pm \vec p_b$. 
The integral in \eq(\ref{eq:cross1}) evaluates to
\beq
 \bigl\langle |\vec p_\CM |^q \bigr\rangle ~=~
   (mT)^{q/2} \, A(q)~,
\label{eq:qresult2}
\eeq
where
\beq 
  A(q) ~\equiv~ \frac{2}{\sqrt{\pi}} \,
    \Gamma\left(\frac{q+3}{2}\right)~.
\label{eq:Adef}
\eeq
Indeed, the Euler gamma-function is formally divergent for $q= -3$, and its 
analytical continuation to smaller values of $q$ is unphysical, as discussed above. 
We note that for $q=2$ we have $A(2)=3/2$ and thus Eq.~(\ref{eq:qresult2}) reduces to
the relation
\beq
    \left\langle |\vec p_\CM|^2 \right\rangle~=~ \frac{3}{2}\, mT~.
\eeq

In terms of $A(q)$, we find that the thermally averaged cross-sections in 
Eqs.~(\ref{eq:cross1}) and (\ref{eq:cross2}) are given by
\beq
  \left\langle \sigma v\right\rangle ~=~ 
   C\,\frac{\left\langle |\vec p_\CM|^q\right\rangle}{m^q} 
    ~=~ C \,\left(\frac{T}{m}\right)^{q/2} A(q)~
\eeq
and 
\beqn
&& \Bigl\langle 
(\KE_a+\KE_b) \sigma v 
\Bigr\rangle\nonumber\\
&&~~~~~~=~
 \frac{C}{m^{q+1}} \left\lbrack
    \langle |\vec p_\CM|^{q+2} \rangle + \langle |\vec p_\CM|^{q} \rangle 
 \langle |\vec p_\CM|^{2} \rangle 
 \right\rbrack\nonumber\\
&&~~~~~~=~
C \,\left(\frac{T}{m}\right)^{q/2} \,T\,
 \biggl[ 
   A(q+2) + A(q)A(2)\biggr]\nonumber\\
 &&~~~~~~=~
C \,\left(\frac{T}{m}\right)^{q/2} \,T\,
 \left( \frac{q+6}{2}\right) 
        A(q)   ~.~~~~~~~
\label{gammeq}
\eeqn
In passing to the final line of Eq.~(\ref{gammeq}) we have utilized the properties 
of the Euler gamma function.  Thus, we find that our pumps are given by
\beqn
  P^{(\rho)}_{M,\gamma} ~&=&~
    \frac{ \rho_M^2}{m} \, C \,\left(\frac{T}{m}\right)^{q/2} \, A(q)\nonumber\\
  P^{(\rho)}_{\KE,\gamma} ~&=&~
    \frac{\rho_M^2 T}{2 m^2} \left(\frac{q+6}{2}\right) 
    C \,\left(\frac{T}{m}\right)^{q/2} \,A(q) ~.~~~
\label{quadratically}
\eeqn
Interestingly, it follows from these relations that
\beq
  P_{\KE,\gamma} ~=~
    \left( 1+\frac{q}{6}\right) \,\frac{\Omega_\KE}{\Omega_M} \,
  P_{M,\gamma} ~.
\label{pumpspushTupwards}
\eeq
This relation between our pumps is essentially a consequence of the thermal distribution 
of momenta we have been assuming for our population of $\phi$ particles --- a distribution 
which essentially enforces a correlation between the momenta (and hence the kinetic energies) 
of these particles and their masses.  We thus find that Eq.~(\ref{pumpbalance}) is true for 
all $q<0$, indicating that our pumps collectively push in the direction of {\it increasing}\/ 
the temperature of our $\phi$-particle gas for all negative $q$. 

We may also intuitively understand why these pump effects on the temperature depend so 
critically on the sign of $q$.  Given the form of the annihilation cross-section in 
Eq.~(\ref{eq:swept_vol_rate}), we see for $q<0$ that it is the more {\it slowly}\/-moving 
$\phi$ particles within our gas that are preferentially annihilated.  The rapid 
re-thermalization of the remaining $\phi$ particles then results in a temperature for 
these particles which is higher than it had previously been.  By contrast, for $q>0$, 
it is the more {\it rapidly}\/-moving $\phi$ particles that are preferentially annihilated.  
Upon re-thermalization, the temperature for the remaining $\phi$ particles is lower 
than it had previously been.

%%%%%%%%%%%%%%%%%%%%%%%%%%%%%%%%%%%%%%%%%%%%%%%%%%%%%%%%%%%%%%%%%%%%%%%%%%%%%%

\section{Dynamical Equations Redux and the Role of Coldness\label{sec:coldness}}

%%%%%%%%%%%%%%%%%%%%%%%%%%%%%%%%%%%%%%%%%%%%%%%%%%%%%%%%%%%%%%%%%%%%%%%%%%%%%%

Given these pumps, we can now construct the differential equations that govern the 
time-evolution of our system within this cosmology.
In principle, there are two quantities whose variations are of interest to us:
$\Omega_M$ and $\Omega_\KE$.  Indeed, as is evident from Eq.~(\ref{KETrelation}), 
the temperature $T$ is simply related to the quotient $\Omega_\KE/\Omega_M$.
However, $\rho_M$ remains as an additional independent quantity within this system 
because the expressions in Eq.~(\ref{quadratically}) for our energy-density pumps 
depend {\it quadratically}\/ on $\rho_M$.  Indeed, although one of these factors 
of $\rho_M$ is converted to $\Omega_M$ when constructing our abundance pumps 
$P_{ij}$ in Eq.~(\ref{eq:abundance_pump}), the other factor of $\rho_M$ remains unconverted. 
Thus, in this system, our differential equations cannot be reduced to abundances alone, 
and the absolute matter energy density $\rho_M$ --- or equivalently the Hubble parameter ---
remains as an additional degree of freedom.  This is a new feature that did not appear 
in any of the previous stasis analyses in Refs.~\cite{Dienes:2021woi,Dienes:2022zgd,Dienes:2023ziv}.
However, this feature arises in the present case as a new consequence of our choice of pump.

We thus have {\it three}\/ independent quantities whose time evolution we wish to study:  
$\Omega_M$, $\Omega_\KE$ (or equivalently $T$), and $\rho_M$.  While we could choose to 
work in terms of these quantities directly, it turns out that our differential equations 
will take their simplest forms if we choose to work in terms of the quantities 
$\Omega_M$, $\Omega_\KE$, and $\Xi$, where
\beq
  \Xi~\equiv~ \frac{T^q \rho_M}{m^{q+4}}~.
\label{eq:DefOfSParam}
\eeq
Because we shall eventually find that our region of interest has $q<0$, we see 
that $\Xi$ increases as $T$ decreases, and {\it vice versa}\/.
We shall therefore refer to $\Xi$ as a {\it coldness}\/ parameter.

Putting all the pieces together, we then find after some algebra that the system of 
equations in Eq.~(\ref{eq:convert3}) can be expressed directly in terms of 
$\Omega_M$, $\Xi$, and $\Omega_\KE$:
\beqn
  \frac{d\Omega_M}{d\calN} ~&=&~ \Omega_M \left[1-\Omega_M 
    + \Omega_\KE- \widehat{C} A(q) \sqrt{ \Xi\Omega_M} \, \right]\nonumber\\
  \frac{d\Xi}{d\calN} ~&=&~  \Xi\left[ -\left(2q+3\right)   
    -\widehat{C} \left(1+\frac{q^2}{6}\right) A(q) \sqrt{ \Xi \Omega_M} 
    \right]~~\nonumber\\
  \frac{d\Omega_\KE}{d\calN} 
    ~&=&~\Omega_\KE \biggl[  -1-\Omega_M + \W_\KE \nonumber\\
    &&~~~~~~~~~~~~~ - \widehat{C} 
    \left( 1 + \frac{q}{6} \right) A(q) \sqrt{ \Xi\Omega_M} \biggr]~ ~~~
\label{Sweqs_expanded}
\eeqn
where we have bundled our time-independent constants together as
\beq
  \widehat C ~\equiv~ \sqrt{\frac{3}{8\pi G}}\, m\, C~
\label{Chatdef}
\end{equation}
and where we have replaced our time variable $t$ with $\calN$, where
\beq
  \calN ~\equiv~ 
  \log\left( \frac{a}{a_0}\right)~.
\eeq
Through the use of the relation $d\calN= Hdt$, we see that $\calN$  is
the number of $e$-folds 
of cosmological expansion which have occurred since an early fiducial time at 
which the scale factor was $a_0$.

%%%%%%%%%%%%%%%%%%%%%%%%%%%%%%%%%%%%%%%%%%%%%%%%%%%%%%%%%%%%%%%%%%%%%%%%%%%%%%

\section{Fixed-Point Solutions and Thermal Stasis\label{sec:fixed_point}}

%%%%%%%%%%%%%%%%%%%%%%%%%%%%%%%%%%%%%%%%%%%%%%%%%%%%%%%%%%%%%%%%%%%%%%%%%%%%%%

Our interest in this paper concerns the fixed-point solutions to 
Eq.~(\ref{Sweqs_expanded}), as such solutions might lead to stasis.  Given the 
dynamical equations in Eq.~(\ref{Sweqs_expanded}), we see that the conditions 
for the existence of a non-trivial fixed-point solution 
$(\barOmega_M,\barXi,\barOmega_\KE)$ with all components non-vanishing are
\beqn
  1-\barOmega_M +\barOmega_\KE ~&=&~ 
    \widehat{C} A(q) \sqrt{ \,\barXi\,\barOmega_M} 
    \nonumber\\
  -( 2q+3) ~&=&~ \widehat{C} \left(1+\frac{q^2}{6}\right) A(q) 
    \sqrt{ \,\barXi \,\barOmega_M}~\nonumber\\
  -1-\barOmega_M +\barOmega_\KE ~&=&~ \widehat{C} 
    \left(1+ \frac{q}{6}\right) A(q) \sqrt{ \,\barXi\,\barOmega_M} ~.~~~~~~~~
\label{Sweqs3}
\eeqn
Since $\widehat C$, $A(q)$, $\barXi$, and $\barOmega_M$ are all non-negative, 
we see from the second line of Eq.~(\ref{Sweqs3}) that no such non-trivial 
fixed-point solutions can possibly exist unless $2q+3<0$, or $q< -3/2$.   

Unfortunately, even with $q< -3/2$, there are  no non-trivial fixed-point 
solutions in which $\barOmega_M$, $\barOmega_\KE$, and $\barXi$ are all non-zero.
However, such a solution does exist in the limit that $\barOmega_\KE \ll 1$ (\ie, 
the limit in which we treat $\barOmega_{\KE}$ as significantly smaller than the other 
abundances, or effectively zero).  This limit is consistent with our original 
assumption that the matter in our theory is non-relativistic.  Within this limit, 
we can disregard the third equation within Eq.~(\ref{Sweqs3}), since the entire right 
side of the corresponding equation in Eq.~(\ref{Sweqs_expanded}) is multiplied by 
$\Omega_\KE$.  We then obtain the non-trivial fixed-point solution given by
\beqn
  \barOmega_M ~&=&~ 1 + \frac{2q+3}{1+q^2/6} \nonumber\\
  \barXi ~&=&~ \frac{1}{\barOmega_M} 
    \left[ \frac{1-\barOmega_M}{\widehat C A(q)} \right]^2 ~.
\label{eq:stasisvalues}
\eeqn
Interestingly, we see that the fixed-point abundance $\barOmega_M$ does not depend 
on the pump prefactor $\widehat C$.  This is ultimately the case because the 
dynamical equations in Eq.~(\ref{Sweqs_expanded}) are invariant under the simultaneous 
transformations $\Xi\to \alpha \Xi$ and $\widehat C\to \widehat C/\sqrt{\alpha}$,
where $\alpha$ is an arbitrary scaling parameter.  This observation is analogous 
to the observation that the quantity $\barOmega_M$ in Ref.~\cite{Dienes:2021woi} 
is independent of $\Gamma_0$, as well as similar observations in 
Refs.~\cite{Dienes:2022zgd,Dienes:2023ziv}.

Note that the restriction $q< -3/2$ implies that $\barOmega_M < 1$.  We likewise find 
that $\barOmega_M > 0$ provided that $q > -6+2\sqrt{3}$.  We shall 
therefore limit our consideration of this fixed-point solution to values of $q$ 
within the range 
\beq
  q_{\rm min} ~<~ q ~<~ q_{\rm max}~
\label{eq:q-range}
\eeq
where
\beqn
  q_{\rm min} ~&=&~ -6 + 2\sqrt{3} ~\approx~ -2.536 \nonumber\\
  q_{\rm max} ~&=&~ -3/2 ~.
\label{qminmaxdef}
\eeqn
Indeed, this is the range within which $0< \barOmega_M < 1$.   Note that the 
$q$-range in Eq.~(\ref{qminmaxdef}) is consistent with the restriction $q>-3$ 
described below Eq.~(\ref{eq:thermalf}).

The existence of such a non-trivial fixed-point solution is only one of the 
requirements that must be satisfied in order for a stasis epoch to arise within this 
system.  We must also require that $\Omega_M$, $\Xi$, and $\Omega_\KE$ dynamically 
approach and remain near this fixed-point solution for an extended time interval, 
regardless of initial conditions.  In other words, we must require that our fixed-point 
solution be a dynamical {\it attractor}\/, so that the system necessarily evolves 
towards this fixed point.  Fortunately, as we shall demonstrate below, the non-trivial 
fixed-point solution in Eq.~(\ref{eq:stasisvalues}) is indeed an attractor for all $q$ 
within the range in Eq.~(\ref{eq:q-range}).    

Before continuing, let us also briefly discuss  the ``trivial'' fixed-point solutions of
Eq.~(\ref{Sweqs_expanded}) --- \ie, solutions wherein $\Omega_M=0$ and/or $\Xi=0$.  
Such solutions are trivial in the sense that their fixed-point behavior for $\Omega_M$ 
and $\Xi$ does not rely on the delicate simultaneous cancellations of the terms within 
each of the corresponding square brackets within the top two lines in Eq.~(\ref{Sweqs_expanded}).  
It turns out that there are three such trivial fixed-point solutions:
\begin{itemize}
\item $\barOmega_M=0$, $\barXi=0$.  This solution is not an attractor, however.
\item $\barOmega_M=0$, $\barXi$ arbitrary.   This is a solution only for $q= -3/2$.  
  However, this solution also fails to be an attractor.
\item $\barOmega_M=1$, $\barXi=0$.  It turns out that this solution is a repellor 
  for $q<-3/2$, but an attractor for $q> -3/2$.  
\end{itemize}
Due to its occasional behavior as an attractor, the third trivial fixed-point solution 
itemized above will also play a role in our analysis of this system.

Thus, to summarize, we find that our system has only one fixed-point solution which 
is also an attractor for any $q> q_{\rm min}$.   For $q< q_{\rm max}$, this attractor 
is the non-trivial fixed-point solution given in Eq.~(\ref{eq:stasisvalues}).   This 
solution will be our primary focus in this paper.   However, for $q> q_{\rm max}$, 
the third trivial fixed-point solution itemized above  becomes our attractor.  

\begin{figure}[t]
  \centering
  \includegraphics[keepaspectratio, width=0.49\textwidth]{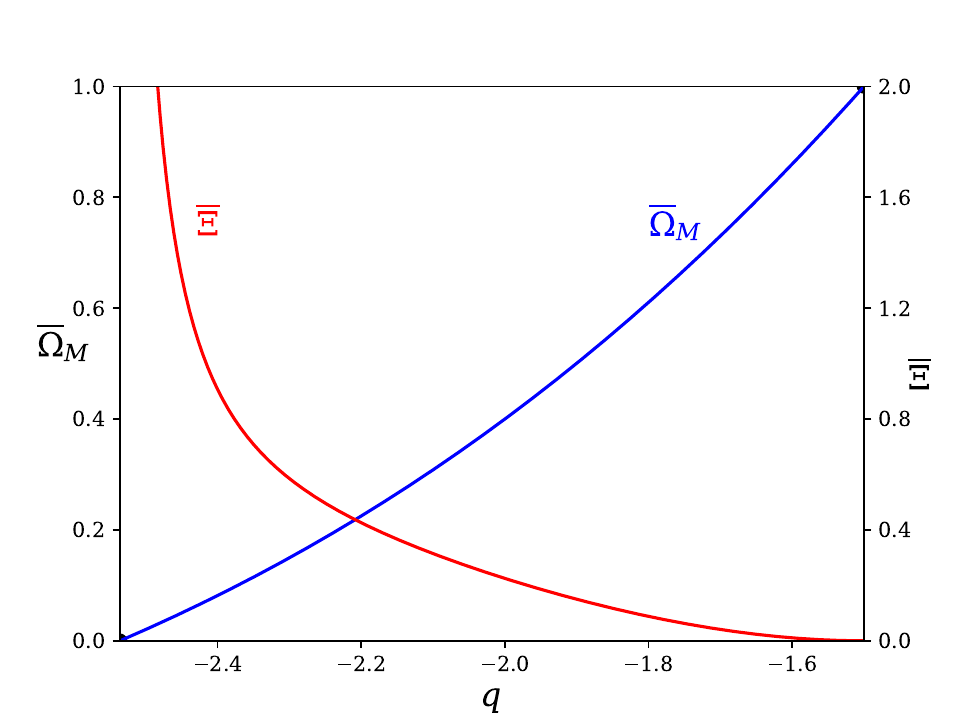}
  \caption{The stasis abundance $\barOmega_M$ (blue) and coldness $\barXi$ (red), 
    plotted as functions of $q$ within the allowed range $q_{\rm min} <q< q_{\rm max}$, with $\widehat{C} = 1$.   
    The abundance curve is plotted relative to the axis along the left edge of the frame, 
    while the coldness is plotted relative to the axis along the right edge.  
    We see that $\barOmega_M$ varies between $0$ and $1$ within this range, as expected, 
    while $\barXi\to \infty$ as $q\to q_{\rm min}$ and  $\barXi\to 0$ as $q\to q_{\rm max}$.
    \label{fig:Wbar_of_q1}}
\end{figure}

In Fig.~\ref{fig:Wbar_of_q1}, we plot the non-trivial fixed-point abundance $\barOmega_M$ 
in Eq.~(\ref{eq:stasisvalues}) as a function of $q$ within the range in Eq.~(\ref{eq:q-range}).    
We also plot the fixed-point coldness $\barXi$ in Eq.~(\ref{eq:stasisvalues}) as a function 
of $q$, taking $\widehat C=1$ as a reference value.  We see that $\barOmega_M$ varies 
between $0$ and $1$ within this range, as expected, while $\barXi$ asymptotes to zero for 
larger $q$ and diverges as $q$ approaches the lower limit of the allowed range.

\begin{figure*}[thb]
  \centering
  \includegraphics[keepaspectratio, width=0.48\textwidth]{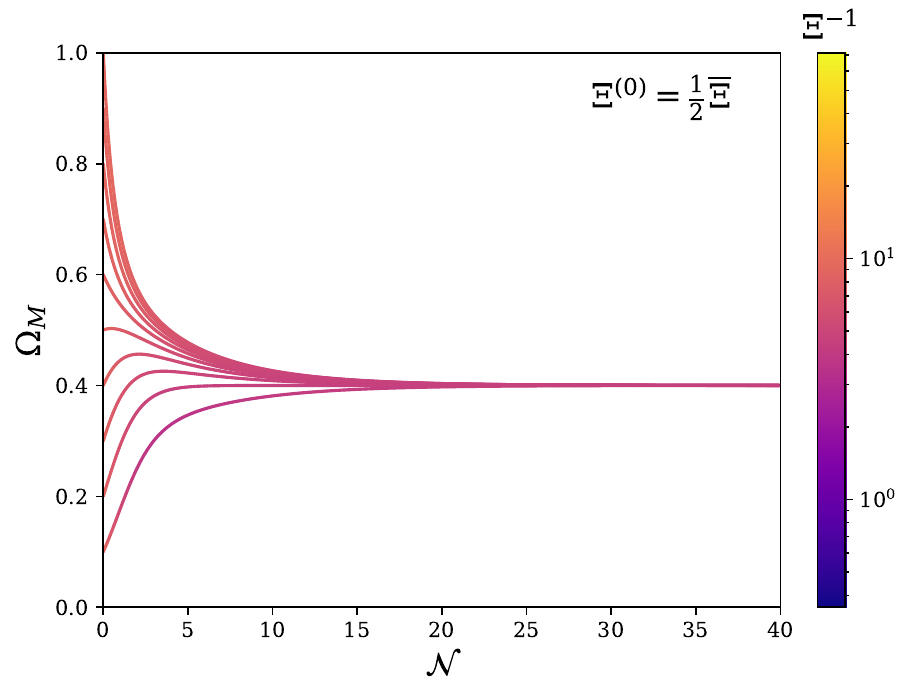}
  \includegraphics[keepaspectratio, width=0.48\textwidth]{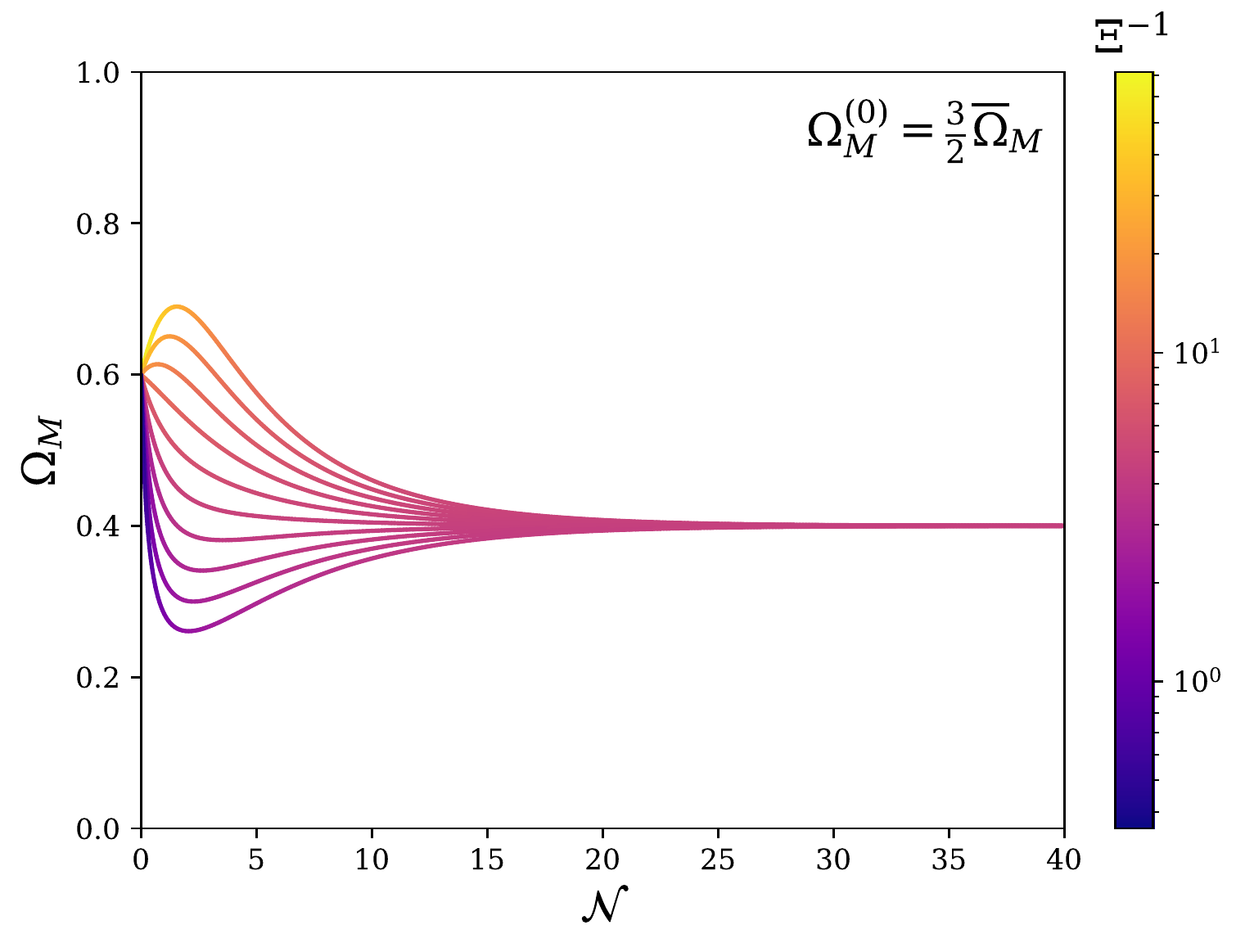}
  \caption{The matter abundance $\Omega_M$, plotted as a function of the number $\calN$ of 
    $e$-folds of cosmological expansion which have occurred since an early fiducial time.  
    At each moment the color of the curve indicates the corresponding value of the coldness 
    $\Xi$.  For all plots, we have taken benchmark values $q= -2$ and $\widehat{C} = 1$
    and simply adjusted either the initial abundance $\Omega_M$ or the initial coldness 
    $\Xi$, as shown in the figures. Indeed, both quantities eventually begin to exhibit 
    stasis, with constant values persisting across many $e$-folds of cosmological 
    expansion.
\label{thermal_stasis_plot}}
\end{figure*}

In Fig.~\ref{thermal_stasis_plot}, we plot  the time-evolution of $\Omega_M$ and $\Xi$ as a 
function of the number $\calN$ of $e$-folds of cosmological expansion which have occurred 
relative to an early fiducial time.   The different curves in each panel correspond to 
taking a variety of different initial conditions. For both panels we have taken $q= -2$ 
and $\widehat{C} = 1$ as benchmark values within the range specified in Eq.~(\ref{eq:q-range}).  
The curves in the left panel have the same initial coldness $\Xi^{(0)}$ but different initial
abundance $\Omega_M^{(0)}$, whereas the curves in the right panel have different initial 
coldness $\Xi^{(0)}$ but the same initial abundance $\Omega_M^{(0)}$.
Most importantly, we observe that in all cases the different curves within each panel 
eventually begin to exhibit  stasis, with an essentially unchanging matter abundance 
$\barOmega_M$ persisting across many $e$-folds of cosmological expansion.  Indeed, 
we find that $\barOmega_M=0.4$ in each panel, which we see from Eq.~(\ref{eq:stasisvalues})
is consistent with our chosen benchmark value $q= -2$. We also note that the {\it colors}\/ of 
these plots (\ie, the corresponding values of the coldness $\Xi$) also evolve towards a 
fixed value $\barXi$ which also persists across many cosmological $e$-folds. These plots also 
provide evidence that our non-trivial fixed-point solution in Eq.~(\ref{eq:stasisvalues}) is 
indeed an attractor, with all curves eventually approaching this fixed-point solution regardless 
of the particular initial conditions assumed.   

%==================
%==================

%=====================================================
%=====================================================

We thus conclude that our system necessarily evolves into an extended stasis epoch, 
regardless of its initial conditions pertaining to either the initial matter (or radiation) 
abundance or the initial coldness.   {\it This is therefore a direct demonstration that the 
stasis phenomenon can be extended into the thermal domain, with a prolonged period of 
stasis experienced not only by the energy abundances but also by thermodynamic 
quantities --- such as coldness --- which involve the temperature of the matter prior to 
annihilation.}

At first glance it might seem that this is not a true thermal stasis because we have 
approximated $\barOmega_\KE\ll 1$ in deriving this non-trivial fixed-point solution.  
Phrased somewhat differently and more precisely, the fixed-point solution 
has $\barOmega_\KE=0$, thus suggesting that the temperature at that point is zero.   
However, while this is true, we must remember that in general our system does not 
spend the many $e$-folds of stasis (such as those shown in Fig.~\ref{fig:Wbar_of_q1}) 
sitting {\it at}\/ the fixed-point solution.   Instead, strictly speaking, this time 
is spent in {\it approaching}\/ the fixed-point solution more and more closely.  
This is true for all of the stasis situations studied in the literature.  Moreover, 
in the present case, we have not only a non-zero $\Omega_\KE$ at all points along 
this evolution, but also a non-zero temperature.   Thus, we have a true thermal situation 
throughout this time period, and our pumps are operating at all points along this trajectory.

This stasis epoch can therefore be characterized as follows:
\begin{itemize}
\item  Our annihilation pumps transfer matter to radiation and serve to counteract the 
tendency of the matter (radiation) abundance to increase (decrease) due to cosmological 
expansion.  In this way, the abundances $\Omega_M$ and $\Omega_\gamma$ are kept 
essentially constant at non-trivial values which depend on $q$ but not on $\widehat C$.

\item  While this is happening, the actual matter energy density $\rho_M$ of 
the $\phi$-particle gas is dropping rapidly due to the combined effects of 
cosmological expansion and $\phi$-particle annihilation.

\item  Likewise, the temperature $T$ of the 
$\phi$-particle gas is also dropping.   This occurs due to cosmological expansion 
[as evident from the first term on the right side of Eq.~(\ref{eq:TempDiffEq})], but 
this effect is partially mitigated by our pumps since $q<0$ [as discussed in
Eq.~(\ref{pumpspushTupwards})].

\item Even though both $\rho_M$ and $T$ are dropping throughout the stasis epoch,
they are dropping in such a balanced way as to hold the coldness 
$\Xi\equiv T^q \rho_M/m^{q+4}$ fixed at a non-trivial value which depends on both 
$q$ and $\widehat C$.

\item The fact that the temperature of the $\phi$-particle gas is dropping during 
the stasis epoch implies that $\Omega_\KE$ is also dropping, thereby justifying 
setting $\Omega_\KE=0$ when deriving our non-trivial limiting fixed-point solution.   
Moreover, the fact that $\Omega_\KE$ quickly becomes negligible during the approach 
to the fixed point is responsible for the fact that we can approximate 
$\Omega_\gamma \approx 1-\Omega_M$ as we approach stasis along this trajectory, 
which is why a constant $\Omega_M$ implies a constant $\Omega_\gamma$ as well.
\end{itemize}

Given this state of affairs, it is natural to wonder how this stasis is able to persist 
across so many $e$-folds of cosmological expansion --- a concern heightened by the fact 
that only a single matter species is involved.   Ultimately, this feature is tied to the 
unique form of our pump.  As we have seen, our $\phi$-particle annihilation pumps for 
both energy density and kinetic-energy density scale as 
\beq   
     P^{(\rho)}~\sim~ \rho_M^2
     ~~~~~\Longrightarrow~~~~~
     P ~\sim~ \rho_M \Omega_M ~.
\label{pumpscaling}
\eeq
Indeed, we see that even our abundance-transferring pumps $P$ have a factor which 
scales with $\rho_M$ rather than with $\Omega_M$.  However, $\rho_M$ is dropping to 
zero throughout our stasis epoch.  {\it Thus, as the  energy density associated with 
our single matter species $\phi$ slowly disappears from our system, the corresponding 
pump also slowly turns off.}\/   Indeed, our pumps always operate in direct 
proportion to the remaining energy density.  Our stasis therefore remains in balance 
even as we proceed further and further out along the tail of the phase-space 
distribution of our slowly vanishing population of $\phi$ particles.  We emphasize 
that this feature did not appear in any previous stasis discussions in the 
literature since the pumps $P^{(\rho)}$ that were utilized in the previous cases 
were all {\it linear}\/ in $\rho_M$, implying that $P\sim \Omega_M\sim {\rm constant}$ 
as far as their dependence on energy densities is concerned.  We thus see that the 
stasis we have derived here has a very different phenomenology and operates 
in a fundamentally different fashion.

Of course, in writing Eq.~(\ref{pumpscaling}) we have disregarded the 
{\it temperature}\/-dependence of our pumps.   From Eq.~(\ref{quadratically}) we see that
\beq
    P_{M,\gamma}~\sim~ T^{q/2}~,~~~~~
    P_{\KE,\gamma} ~\sim~ T^{q/2+1}~.
\eeq
Thus, while $P_{M,\gamma}$ always scales inversely with $T$ 
as the temperature of our $\phi$-particle gas drops,
we see that $P_{\KE,\gamma}$ scales inversely with $T$ only for $q> -2$.  Indeed, for smaller $q$ 
we find that both $P_{\KE,\gamma}$ and the temperature $T$ drop together --- an effect that lies 
beyond the energy-density scaling effect discussed above and which tends to further suppress the 
activity of this pump as our $\phi$-particle gas cools and dissipates.

%%%%%%%%%%%%%%%%%%%%%%%%%%%%%%%%%%%%%%%%%%%%%%%%%%%%%%%%%%%%%%%%%%%%%%%%%%%%%%

\section{Thermal Stasis as a Global Attractor\label{sec:attractor}}

%%%%%%%%%%%%%%%%%%%%%%%%%%%%%%%%%%%%%%%%%%%%%%%%%%%%%%%%%%%%%%%%%%%%%%%%%%%%%%

\begin{figure}[t!]
  \centering
  \includegraphics[keepaspectratio, width=0.49\textwidth]{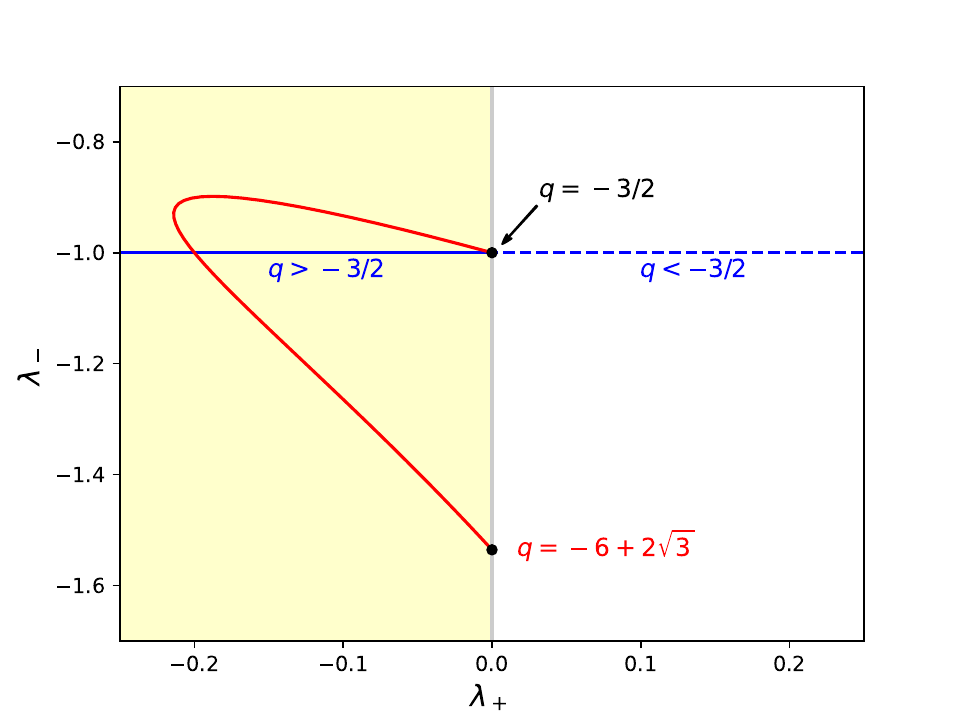}
  \caption{Parametric plot of the Jacobian eigenvalues $(\lambda_+,\lambda_-)$ 
    as the parameter $q$ is varied.  In all cases these eigenvalues are evaluated 
    for a given fixed-point solution.  The red curve indicates the combinations
    of these eigenvalues associated with the non-trivial fixed point in 
    Eq.~(\ref{eq:stasisvalues}), while the blue horizontal line indicates the eigenvalues 
    associated with the trivial fixed point described in third bullet below 
    Eq.~(\ref{qminmaxdef}).  Thus, we see that not only are $\barOmega_M$ and 
    $\barXi$ independent of $q$ for this trivial fixed point, but $\lambda_-$ is as 
    well.  As discussed in the text, the non-trivial fixed 
    point is an attractor throughout the entirety of the range 
    $-6+2\sqrt{3}\leq q \leq -3/2$, while the trivial fixed point is an 
    attractor only for $q> -3/2$.  Note that these fixed points are attractors only 
    when both eigenvalues $\lambda_\pm$ are negative and hence lie within the 
    yellow shaded region.   
    The corresponding lines are then shown as solid rather than dashed.  
    Interestingly, the red and blue curves intersect precisely at $q= -3/2$, 
    where they both correspond to the solution $(\barOmega_M,\barXi)=(0,1)$.   
    The fact that the red and blue curves intersect at only one point ensures 
    that there is never more than one attractor solution  as $q$ crosses the 
    $q= -3/2$ boundary.  Thus, as $q$ increases from $q= -6+2\sqrt{3}$, our 
    attractor solution follows the red curve until reaching the $q=-3/2$ point 
    and then switches to the solid blue line for $q> -3/2$.
\label{fig:temp:lambdas}}
\end{figure}

It is straightforward to demonstrate that the non-trivial fixed-point solutions 
for $(\barOmega_M,\barXi)$ 
in Eq.~(\ref{eq:stasisvalues}) are also global attractors.  We have already seen 
evidence of this attractor behavior in Fig.~\ref{thermal_stasis_plot}.  As a first 
step, we note that for a given value of $q$, we may assess whether the 
corresponding fixed-point solution is a {\it local}\/ attractor by
evaluating the $2\times 2$ Jacobian matrix for our dynamical equations 
for $d\Omega_M/d\calN$ and $d\Xi/d\calN$ and then examining the signs of its two 
eigenvalues $\lambda_\pm$ at that fixed point.  In Fig.~\ref{fig:temp:lambdas}, we 
present a parametric plot of these eigenvalues as $q$ is varied within the 
range $q_{\rm min}<q< q_{\rm max}$ for which our non-trivial fixed-point solution exists.
These eigenvalues trace out the red solid curve.  Within this $q$-range, we see that both 
eigenvalues lie within the yellow-shaded region wherein both $\lambda_+$ and $\lambda_-$ 
are negative.  Thus, we may conclude that across this entire range of $q$, the 
non-trivial fixed-point solution within this range is indeed a local attractor.

%==========================================
%==============================================================

Within Fig.~\ref{fig:temp:lambdas} we have also plotted the eigenvalues
associated with the {\it trivial}\/ fixed-point solution indicated in the third bullet 
in the paragraph below Eq.~(\ref{qminmaxdef}).  These eigenvalues fill out the blue 
horizontal line.  When $q<q_{\rm max}=-3/2$, we see that $\lambda_+$ is positive, 
telling us that this solution is not an attractor for such values of $q$.   However, 
as $q$ increases beyond $q_{\rm max}$, we pass across into the yellow-shaded region 
within which both eigenvalues are negative and within which this trivial fixed-point 
becomes an attractor.  Interestingly, the blue line associated with our trivial 
fixed-point solution intersects the red curve associated with our non-trivial 
fixed-point solution precisely at $q= q_{\rm max}=-3/2$.  Indeed, as $q$ increases 
through the point $q = -3/2$, our non-trivial fixed-point solution {\it ceases}\/ to 
be an attractor and our trivial fixed-point solution {\it becomes}\/ an attractor.  
Thus $q=-3/2$ marks a continuous boundary between these two different solutions.   
Indeed, even the corresponding solutions for $\barOmega_M$ and $\barXi$ are continuous 
across this boundary.

\begin{figure*} 
%  [t!]
  \centering
  \includegraphics[keepaspectratio, width=0.8\textwidth]{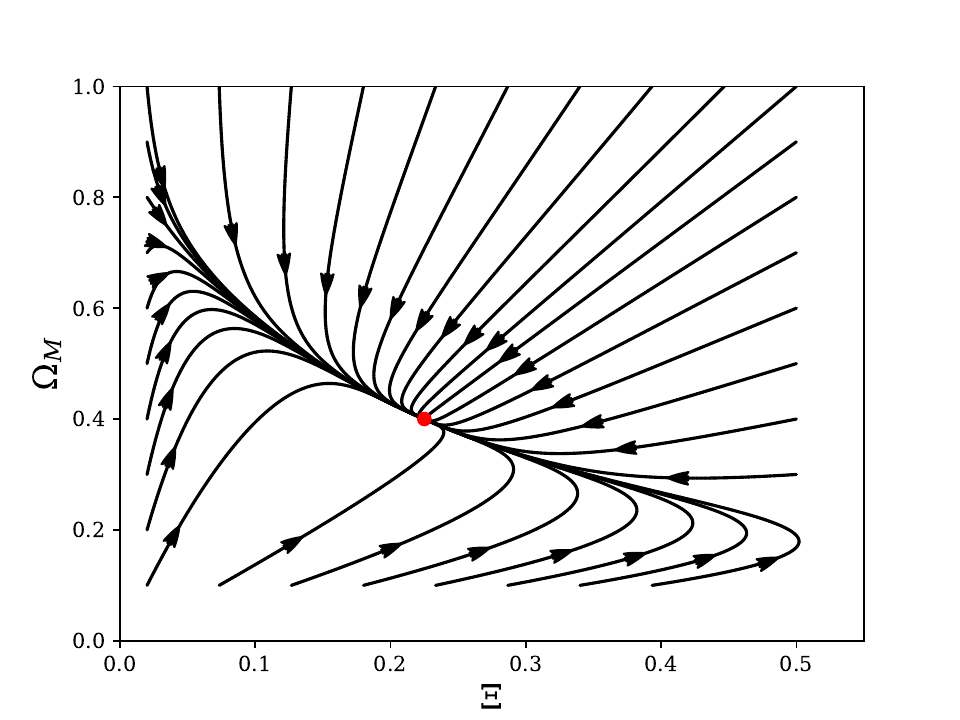}
  \caption{
    Attractor behavior for our thermal stasis when $q$ lies within the allowed stasis 
    range in Eq.~(\ref{eq:q-range}).  These trajectories within the $(\Xi,\Omega_M)$-plane 
    correspond to the time evolution of the cosmological system described in  
    Eq.~(\ref{Sweqs_expanded}),  with chosen benchmark values $q= -2$ and $\widehat C=1$.
    The central red dot towards which all flow lines tend is our corresponding stasis 
    solution with $(\barXi,\barOmega_M)=(9/40,2/5)$.   We note that this plot was made 
    assuming a relatively small initial value for $\Omega_\KE$, but similar plots arise 
    regardless of the initial value of $\Omega_\KE$.
\label{thermal_stasis_plot_2d2}}
%===============================================================
%============================
  \centering
  \includegraphics[keepaspectratio, width=0.48\textwidth]{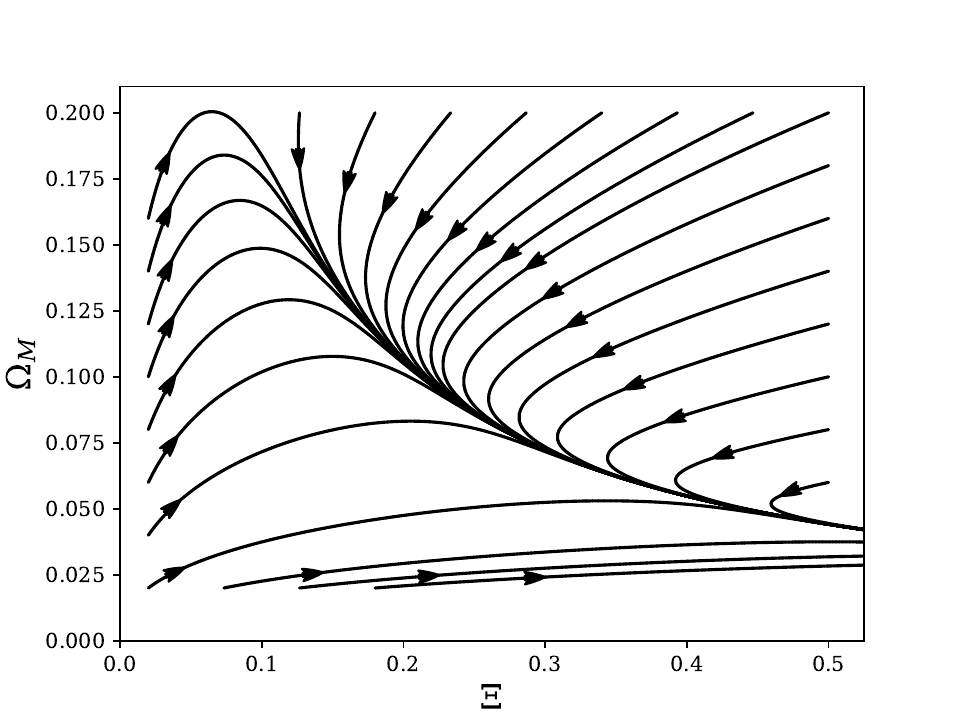}
  \includegraphics[keepaspectratio, width=0.48\textwidth]{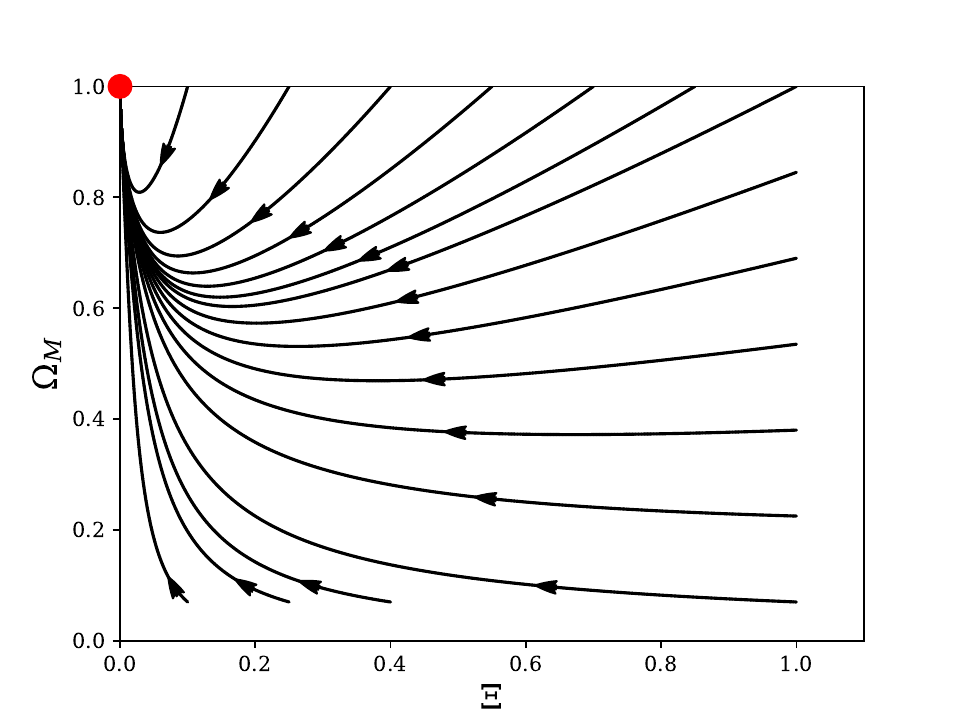}
  \caption{Time evolution of our cosmology when $q$ is outside the stasis range 
    $q_{\rm min} < q < q_{\rm max}$ in Eq.~(\ref{eq:q-range}), with either $q< q_{\rm min}$ 
    (left panel) or $q> q_{\rm max}$ (right panel).  The left and right panels shown 
    correspond to the benchmark values $q= -2.7$ and $q= -1.3$ respectively, with 
    $\widehat C=1$.  Note that $\Xi$ can either increase or decrease if $q< q_{\rm max}$, 
    but can only decrease if $q> q_{\rm max}$.   For $q< q_{\rm min}$, our system has no 
    fixed point, and all of the flows proceed toward small $\Omega_M$ but ever-increasing 
    coldness $\Xi$.  By contrast, for $q>q_{\rm max}$, our system is attracted to the trivial 
    fixed point at $(\barXi,\barOmega_M)=(0,1)$.
    \label{thermal_runaway_plot_2d}}
\end{figure*}

Since the Jacobian analysis we have presented above characterizes only the {\it local}\/ 
behavior of our system --- \ie, its behavior within the vicinity of the fixed 
point --- it still remains to be seen whether these fixed points are in fact 
{\it global}\/ attractors toward which our system dynamically flows regardless of 
its initial conditions.  However, within the $q$-ranges we have been discussing, 
it turns out that this is indeed the case.
We can be verify this by examining the trajectory along which the system 
described in Eq.~(\ref{Sweqs_expanded}) dynamically evolves in the 
$(\Omega_M,\Xi)$-plane, given an arbitrary initial configuration within
that plane.  In Fig.~\ref{thermal_stasis_plot_2d2} we show a number of such trajectories 
for the benchmark value $q = -2$.  We find that regardless of the initial 
location of our system within this plane, our system is inevitably drawn toward our 
non-trivial fixed-point solution, indicating that this fixed point is not only an attractor but also global rather than merely local.

It is also instructive to examine what happens when we choose $q$ {\it outside}\/ 
the range specified in Eq.~(\ref{eq:q-range}).  The behavior of our system ultimately  
depends on whether $q < q_{\rm min}$ or $q> q_{\rm max}$.  Both behaviors are shown in
Fig.~\ref{thermal_runaway_plot_2d}.  In the left plot, we consider a situation with 
$q< q_{\rm min}$.  In this case, our flow lines all tend towards small values of 
$\Omega_M$ with ever-increasing values of $\Xi$.  Indeed, no fixed-point solution 
is ever reached.  By contrast, for $q> q_{\rm max}$, all of our flow lines are pulled 
toward the {\it trivial}\/ fixed point at $(\barXi,\barOmega_M)=(0,1)$, as anticipated.   
Thus even our trivial fixed 
point is a global attractor for $q> -3/2$.

%============================

%%%%%%%%%%%%%%%%%%%%%%%%%%%%%%%%%%%%%%%%%%%%%%%%%%%%%%%%%%%%%%%%%%%%%%%%%%%%%%

\section{A Particle-Physics Mechanism for Achieving Thermal Stasis\label{sec:mechanism_q_neg_2}}

%%%%%%%%%%%%%%%%%%%%%%%%%%%%%%%%%%%%%%%%%%%%%%%%%%%%%%%%%%%%%%%%%%%%%%%%%%%%%%

In the previous section, we demonstrated that an extended epoch of matter/radiation stasis 
can emerge in a thermal context, and with only a single matter species $\phi$, provided not only that 
this species annihilate into radiation via a $2\to 2$ process of the form $\phi\phi\to \chi\chi$, 
but also that the thermally averaged cross-section for this annihilation process take the form 
given in Eq.~(\ref{eq:swept_vol_rate}) with $q$ within the range in Eq.~(\ref{eq:q-range}).~  
In this section, we outline how these conditions can be realized in a particle-physics
context.  

As one might imagine, the principal challenge in constructing a particle-physics model 
of thermal stasis along these lines is to ensure that the swept-volume rate $\sigma v$ for 
matter-particle annihilation takes the form in Eq.~(\ref{eq:swept_vol_rate}) with a
value of $q$ within the desired range in Eq.~(\ref{eq:q-range}).  
Indeed, as we shall discuss below, typical particle-physics models that one can construct 
along these lines lead to values of $q$ which are either above or below this range.  
Such models therefore do not yield a stasis epoch.  However, as we shall now demonstrate, 
there do exist particle-physics mechanisms which lead to suitable values of $q$. 
In what follows, we describe one such mechanism --- one which leads specifically to a 
value $q=-2$ for this parameter.  This mechanism rests on fundamental ideas in quantum 
field theory and can be realized within a broad variety of particle-physics
contexts.  We shall therefore keep the following discussion as model-independent as possible.  

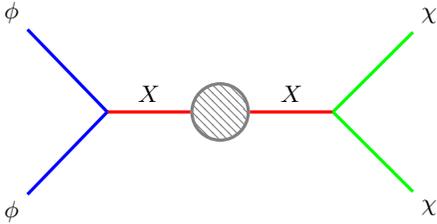
\begin{figure}[H]
    \centering
    \begin{tikzpicture}
    \begin{feynman}
        \node[ blob,draw=gray,pattern color=gray,very thick] (loop);
        \vertex [left=of loop] (mx);
        \vertex [right=of loop] (gx);
        \vertex [below left=of mx] (i1) {\(\phi\)};
        \vertex [above left=of mx] (i2) {\(\phi\)};
        \vertex [above right=of gx] (f1) {\(\chi\)};
        \vertex [below right=of gx] (f2) {\(\chi\)};
        \diagram* {
        (mx) -- [red, plain, very thick, edge label=\(\textcolor{black}{X}\)] (loop) 
          -- [red, plain, very thick, edge label=\(\textcolor{black}{X}\)] (gx),
        (i1) -- [blue, plain, very thick] (mx) -- [blue, plain, very thick] (i2),
        (f1) -- [green, plain, very thick] (gx) -- [green, plain, very thick] (f2)
        };
    \end{feynman}
 \end{tikzpicture}
    \caption{Feynman diagram for the $s$-channel annihilation process through which two $\phi$ 
    particles annihilate into radiation through a mediator $X$, complete with a one-loop radiative 
    correction for $m_X$.~  Here the $\phi$ particles constitute the matter in our model and are 
    indicated in blue, while the resulting radiation is indicated in green and the mediator 
    particle $X$ is indicated in red.  The one-loop correction for the propagator (shown in gray) 
    comprises loops of matter or radiation. 
    \label{pump_diagram}}
\end{figure}

We focus on the case in which the annihilation process $\phi\phi \to \chi\chi$ 
proceeds primarily through an $s$-channel mediator $X$, as illustrated in 
Fig.~\ref{pump_diagram}.  For concreteness, we take $\phi$, $\chi$, and $X$ all to be 
scalar fields in what follows, though we note that this mechanism can be realized for certain 
other combinations of spin assignments as well.  We shall let $m_X$ denote the mass of $X$ 
and continue to let $m$ denote the mass of $\phi$.~  In the CM frame, the incoming $\phi$ 
particles have the four-momenta
\beqn
    p_1 ~&=&~ (E, \vec p_\CM )\nonumber\\
    p_2 ~&=&~ (E, -\vec p_\CM)
\eeqn
where $\pm \vec p_\CM$ are the three-momenta of our 
incoming particles as seen in the CM frame and where we assume that each of these 
incoming particles is on shell.  Thus, for an $s$-channnel propagator in the 
CM frame, the total four-momentum flowing through the $X$ particle is 
\beq
  p_X ~\equiv~ p_1 + p_2 ~=~ (2E, \vec 0)~,
\eeq
from which we deduce that 
\beq 
     p_X^2 ~=~ 4 E^2 ~=~ 4\left(|\vec p_\CM|^2  + m^2\right) ~.
\label{intermed}
\eeq
The tree-level propagator is then given by
\beqn
\Delta(p_X) ~&\sim& ~ \frac{i}{p_X^2 - m_X^2} \nonumber\\
  &=&~ \frac{i}{4(|\vec{p}_\CM|^2 + m^2)  - m_X^2}\nonumber\\
  &=&~ \frac{i}{4 |\vec{p}_\CM|^2 -\mu^2 }~,
\eeqn
where
\beq
   \mu^2 ~\equiv ~ m_X^2 - 4m^2 ~.
\label{eq:mudef}
\eeq
It should be noted that we are {\it not}\/ assuming that $m_X=2 m$ (for which we 
would have $\mu=0$), and thus $\mu$ appears as a new free parameter.  In general, 
we see from Eq.~(\ref{eq:mudef}) that $\mu^2$ can have either sign.

%========= BEGIN FIGURE ========
%========= END FIGURE ========

If $|\mu^2|\ll 4|\vec p_\CM|^2$, then $\Delta \sim 1/ |\vec p_\CM|^2$, which leads 
to $q= -4$.  By contrast, if $|\mu^2|\gg 4|\vec p_\CM|^2$, we find that $\Delta$ becomes 
independent of  $|\vec p_\CM|$ and thus $q=0$.  In all other cases, $q$ will vary between 
$-4$ and $0$.  However, we will never have a non-zero range of $|\vec{p}_\CM|$ values for which 
$q$ is effectively constant and lies within the desired stasis range $-6+2\sqrt{3}<q< -3/2$.

We now investigate what happens when we include the one-loop radiative correction to 
the $X$ propagator.  In particular, we shall let $\Pi_X(p_X^2)$ denote the one-loop 
contribution to the self-energy of the mediator, which corresponds to the one-particle 
irreducible (1PI) ``bubble'' within the $X$ propagator in Fig.~\ref{pump_diagram}.  
Our propagator will then take the form
\beq
  \Delta(p_X) ~\sim~ \frac{i}{p_X^2 - m_X^2 + \Pi_X(p_X^2)}~.
  \label{eq:prop_raw}
\eeq
Despite the fact that $\Pi_X(p_X^2)$ is a one-loop radiative correction, 
we shall find this term can nevertheless become dominant if we are near a 
resonance in the propagator.

The mediator self-energy $\Pi_X(p_X^2)$ depends on the specific 
model under study and is in general a complex-valued function.  Since the real part of 
$\Pi_X(p_X^2)$ can be absorbed by a redefinition of $m_X$, the $X$-propagator 
may be written in the form
\beq
  \Delta(p_X) ~\sim~ 
    \frac{i}{p_X^2 - m_X^2 + i m_X \widetilde{\Gamma}_X (p_X^2) }~,
\label{eq:DeltaFull}
\eeq
where we have defined
\beq
  \widetilde{\Gamma}_X (p_X^2) ~\equiv~\frac{1}{m_X} \Im \,  \Pi_X(p_X^2) ~.
\label{optical}
\eeq
In what follows, we shall refer to this quantity as the ``off-shell width'' of the 
mediator.  Despite this choice of terminology, however, we emphasize that 
$\widetilde{\Gamma}_X(p_X^2)$ is in general a non-trivial function of $p_X^2$, 
and is {\it not}\/ in general equal to the proper decay width of $X$.

For any model which gives rise to an $s$-channel annihilation diagram 
of the form depicted in Fig.~\ref{pump_diagram}, $\widetilde{\Gamma}(p_X^2)$ receives
contributions from two processes: one with a pair of $\chi$ particles running in the loop
and one with a pair of $\phi$ particles running in the loop.  We shall 
denote these contributions $\widetilde{\Gamma}_{X\to \chi\chi}(p_X^2)$ and
$\widetilde{\Gamma}_{X\to \phi\phi}(p_X^2)$, respectively.  
Thus, in general, we have 
\beq
   \widetilde{\Gamma}_X ~=~ 
   \widetilde{\Gamma}_{X\to \chi\chi} + \widetilde{\Gamma}_{X\to \phi\phi}~.
\label{offshelldecaywidths}
\eeq  

Since our regime of interest for stasis is that in which our 
population of $\phi$ particles is highly non-relativistic, we are particularly 
interested in how $\widetilde{\Gamma}_X(p_X^2)$ behaves as a function of 
$|\vec{p}_\CM|$ when $|\vec{p}_\CM|$ is small.  Within this regime, we may obtain a
reliable approximation for $\widetilde{\Gamma}_X(p_X^2)$ by expanding this 
function as a Taylor series in  $|\vec{p}_\CM|$ around $|\vec{p}_\CM| = 0$ and retaining only
the leading terms.  With only minor assumptions about the model, it can be shown that 
the $|\vec{p}_\CM|$-independent term and the term linear in $|\vec{p}_\CM|$ are both necessarily
non-zero.  Thus, within our regime of interest, $\widetilde{\Gamma}_X$ is well 
approximated by an expression of the form
\beq 
   \widetilde{\Gamma}_X ~\approx~ c_0 + c_1 |\vec p_{\CM}|~
\label{eq:TildeGammaX_expansion}
\eeq  
where $c_0$ and $c_1$ are arbitrary non-zero, model-dependent constants.

Substituting this expression into Eq.~(\ref{eq:DeltaFull}), we obtain
\beqn
\Delta(p_X) ~&\sim&~
    \frac{i}{
    4|\vec p_\CM|^2 - \mu^2 + i m_X 
    \left( c_0 + c_1 |\vec p_\CM| \right)} 
\nonumber\\
~&= &~
   \frac{i}{m^2} 
  \left[ \widetilde{\alpha}+i\beta
  \frac{|\vec p_{\CM}|}{m} + 4
   \frac{|\vec p_{\CM}|^2}{m^2}\,\right]^{-1} ~~~ ~~~
\nonumber\\
 ~&=&~ 
 \frac{i}{\alpha m^2}\, \frac{1}{e^{i\theta}+ ixy + x^2}~,
 \label{eq:propag}
\eeqn
where in passing from the second to the third line we have defined 
\beq
  \widetilde \alpha ~=~ -\frac{\mu^2}{m^2} 
         + \frac{i c_0 m_X}{m^2}~,~~~~
 \beta ~=~ c_1 \frac{m_X}{m}~,
\label{eq:abcdef}
\eeq
and where in passing from the third to the fourth line we have expressed
the complex coefficient $\widetilde{\alpha}$ in terms of its modulus $\alpha$ 
and complex phase $\theta$, whence
\beq
  x ~=~ \sqrt{\frac{4}{\alpha}}\, \frac{|\vec p_{\CM}|}{m}~,~~~~
  y ~=~ \frac{\beta}{\sqrt{4\alpha}}~.
\label{eq:xydef}
\eeq 

Our original question was to determine the circumstances under which the propagator 
$\Delta$ might scale as $|p_\CM|^{-1}$, since this would lead to amplitudes with the 
desired scaling with $q= -2$.   Given the above results, we are now in a 
position to answer this question.  Defining $f(x)\equiv 1/| e^{i\theta} + ixy + x^2|$, 
we find
\begin{itemize}
    \item $f(x) \sim {\rm constant}$ for $x\ll 1/y$;
    \item $f(x) \sim 1/x$ for $1/y \ll x \ll y$;  and
    \item $f(x) \sim 1/x^2$ for $x\gg y$.
\end{itemize}
Indeed, these results hold for all $y$ and all $\theta$.  We thus see that we can indeed 
achieve the desired inverse linear scaling for our propagator across a significant interval 
of momenta $|\vec p_\CM|$ so long as
\beq
  y~\gg~ 1~,~~~~ 1/y ~\ll~ x ~\ll~ y ~.
\label{linear}
\eeq
However, as discussed previously, we also know that our incoming $\phi$ particles must be 
non-relativistic (so that they can be interpreted as matter rather than radiation).  This 
then implies the additional constraint 
\beq
  x ~\ll~ \sqrt{4/\alpha}~.
  \label{nonrel}
\eeq

The cosmological implications of the criterion in Eq.~(\ref{linear}) for stasis
can be understood as follows.
Let us assume, as we did in Sect.~\ref{sec:preliminaries}, that additional scattering 
processes in the theory serve to maintain kinetic equilibrium among the population
of $\phi$ particles, and that this population of particles can be characterized by 
a temperature $T$.  At values of $T$ for which the phase-space integrals in 
$\langle \sigma v\rangle$ and $\langle ({\rm KE}_a+{\rm KE}_b)\sigma v\rangle$ are 
each dominated by the contribution from particles with $|\vec{p}_{\rm CM}|$ within 
the range specified in Eq.~(\ref{eq:allowedp}), these thermal averages are 
well approximated by the expressions in Eq.~(\ref{gammeq}) with $q = -2$.  Thus,
at such temperatures, the universe evolves toward stasis.
By contrast, at values of $T$ for which these phase-space integrals receive sizable 
contributions from particles with $|\vec{p}_{\rm CM}|$ outside this range, 
$\langle \sigma v\rangle$ and $\langle ({\rm KE}_a+{\rm KE}_b)\sigma v\rangle$ do 
not exhibit the appropriate scaling behavior with $T$ and the system does not evolve
toward stasis.

In order to obtain an estimate of the range of $T$ within which our system is indeed 
attracted toward stasis, we can associate a given value of $T$ with a ``typical'' 
value of $|\vec p_\CM|$ --- \ie, a value near the peak in the Maxwell-Boltzmann
distribution --- via the relation
\begin{equation}
  T ~\sim~ |\vec p_{\CM}|^2/m~,
  \label{eq:emp_sim_from_pcm}
\end{equation}
where in making this identification we have disregarded ${\cal O}(1)$ factors.
Since the constraint in Eq.~(\ref{linear}), expressed in in terms of 
$|\vec{p}_{\rm CM}|$, is
\beq
  \frac{\alpha}{\beta} m ~\ll~ |\vec p_\CM |
    ~\ll~ \frac{\beta}{4} m~,
  \label{eq:allowedp}
\eeq
it therefore follows that the temperature window wherein the stasis attractor
is the solution to the coupled evolution equations in Eq.~(\ref{eq:convert3}) is
\begin{equation}
  \Tmin ~\lsim ~ T ~\lsim~ \Tmax~,
\label{eq:Trange}
\end{equation}
where we have defined
\begin{eqnarray} 
  \Tmax ~&\equiv &~ \frac{\alpha^2}{\beta^2} m
    \nonumber\\
  \Tmin ~&\equiv &~ \frac{\beta^2}{16} m~.
  \label{eq:T_bounds_propagator}
\end{eqnarray}

As long as $T$ remains within the rough window in Eq.~(\ref{eq:Trange}), the universe 
continues to evolve toward stasis.  However, as time goes on and the temperature
of the $\phi$-particle gas decreases, eventually $T$ drops below $\Tmin$ and the 
stasis attractor ceases to be the solution to the coupled evolution equations in 
Eq.~(\ref{eq:convert3}).  Indeed, when $T$ falls below this threshold,  
$\langle \sigma v\rangle$ no longer increases as the kinetic-energy density 
of the $\phi$ particles decreases.  As a result, $\phi\phi\to\chi\chi$ 
annihilation becomes inefficient and $\Omega_M$ increases until the
rest-mass-energy density of the $\phi$-particle gas dominates the total 
energy density of the universe.  This, then, is the 
manner in which stasis naturally ends in models which make use of this mechanism, 
with the universe subsequently becoming dominated by massive matter.

Of course, any particle-physics model which gives rise to an annihilation process of the 
form illustrated in Fig.~\ref{pump_diagram} necessarily also gives rise to a number of
scattering processes which can in principle impact the cosmological dynamics of our system.
Two such processes are $\phi\phi\to\phi\phi$ scattering and $\chi\chi\to\chi\chi$ scattering,
which, if efficient, can serve to establish and maintain kinetic equilibrium among the 
cosmological populations of $\phi$ and $\chi$ particles, respectively.  However, it turns out that 
within the particle-physics models that are capable of achieving thermal stasis via the mechanism described above, 
$\phi\phi\to\phi\phi$ scattering is typically efficient whereas $\chi\chi\to\chi\chi$ scattering typically is not.    
This is because this mechanism can only give rise to a stasis epoch with a significant duration 
when $T_{\rm min} \ll T_{\rm max}$ and the condition in Eq.~(\ref{eq:Trange}) is therefore satisfied 
across a broad range of $T$.  This typically requires that the $\phi$ particles couple to the 
mediator $X$ with a far greater strength than do the $\chi$ particles~\cite{Barber_toappear_model}.  
As a result, the population of $\phi$ particles remains in kinetic equilibrium in stasis models of this 
sort, whereas the population of $\chi$ particles produced by $\phi\phi\to\chi\chi$ annihilation typically never 
attains kinetic equilibrium and therefore possesses a highly non-thermal phase-space distribution.

A third process which necessarily arises in particle-physics models of this sort
is $\chi\phi\to\chi\phi$ scattering --- a process which facilitates the transfer of kinetic 
energy from radiation to matter.  Scattering processes of this sort can potentially 
have a significant impact on the manner in which the temperature of a cosmological population of 
non-relativistic particles evolves over time in scenarios with non-standard expansion histories.  
For example, the manner in which a population of cold dark-matter particles kinetically decouples from the 
radiation bath during an early matter-dominated era (EMDE) can be very different from the manner in 
which such a population of particles decouples during a radiation-dominated era.  
In particular, due to the continuous injection of radiation from the heavy decaying particles which 
dominate the energy density of the universe during the EMDE, this decoupling process can include a 
lengthy {\it quasi-decoupling}\/ phase wherein the temperature of the matter particles evolves 
in a non-standard manner~\cite{Waldstein:2016blt}.  One might therefore worry that
a process like $\chi\phi\to\chi\phi$ scattering could alter the manner in which $T$ evolves with
time in our thermal-stasis scenario such that the system does not in fact gives rise to stasis at all. 

The impact that $\chi\phi\to\chi\phi$ scattering has on the evolution of $T$ within any 
particular particle-physics model in which our mechanism for achieving thermal stasis is realized 
depends sensitively on the details of the model.  In what follows, we summarize the qualitative 
manner in which this process affects the cosmological dynamics.  A more detailed, quantitative 
analysis, performed within the context of a minimal such model, is provided in  
Ref.~\cite{Barber_toappear_model}.  It turns out that $\chi\phi\to\chi\phi$ scattering modifies 
the evolution of $T$ in thermal-stasis scenarios of this sort by giving rise to an $\Omega_M$-dependent 
correction to the coefficient of the first term in Eq.~(\ref{eq:dTdtequation}).  In particular, in 
the presence of such scattering, this equation is modified to
\begin{eqnarray}
  \frac{dT}{dt} &~=~&
    - \left[2- \frac{(1-\Omega_M)^2}{3\Omega_M}\,\epsilon\right]  H T 
    \nonumber \\ & & ~~~~~ -\,\frac{2m}{3\Omega_M} 
    \left( P_{\KE,\gamma} 
    - \frac{\Omega_\KE}{\Omega_M}
    \,P_{M,\gamma} \right)~,~~~~~~
  \label{eq:dTdtequationMod}
\end{eqnarray}
where $\epsilon$ is a constant factor which depends on the couplings involved.  This 
correction term leads in turn to a correction term in the dynamical equation 
for $\Xi$ in Eq.~(\ref{Sweqs_expanded}).  

Since $\chi\phi\to \chi\phi$ scattering proceeds through a $t$-channel process and therefore
does not receive the sizable resonant enhancement that $\phi\phi\to\chi\chi$ annihilation
receives, it is often the case that $\epsilon$ is extremely small.  Indeed, one finds
that the correction term in Eq.~(\ref{eq:dTdtequationMod}) is often negligible within
the parameter-space regions of interest in thermal-stasis models of this sort and can therefore
be ignored.  Interestingly, however, it also turns out that the structure of this correction 
term is such that stasis emerges in this system and remains a global dynamical attractor 
even when $\epsilon$ is sufficiently large that this term cannot be neglected.  Indeed,
this term simply leads to a modification of the resulting stasis values $\barOmega_M$ and 
$\barXi$.  Thus, while $\chi\phi\to\chi\phi$ scattering enriches the cosmological 
dynamics in such model realizations of thermal stasis,
it does not impact our qualitative results.

%%%%%%%%%%%%%%%%%%%%%%%%%%%%%%%%%%%%%%%%%%%%%%%%%%%%%%%%%%%%%%%%%%%%%%%%%%%%%%

\section{Conclusions and Directions for Future Research\label{sec:conclusions}}

%%%%%%%%%%%%%%%%%%%%%%%%%%%%%%%%%%%%%%%%%%%%%%%%%%%%%%%%%%%%%%%%%%%%%%%%%%%%%%

In all of its realizations, cosmological stasis is ultimately a consequence of
processes which transfer energy density from one cosmological component to
another in a way which counteracts the effects of Hubble expansion.  
A variety of processes, including particle 
decay~\cite{Dienes:2021woi,Dienes:2023ziv}, Hawking 
radiation~\cite{Barrow:1991dn,Dienes:2022zgd}, and the overdamped/underdamped 
transition of homogeneous scalar-field zero-modes~\cite{Dienes:2024wnu} give 
rise to pumps which can accomplish this feat.  However, the pump rate associated with 
each of these processes depends only on the intrinsic properties 
(\eg, masses or decay widths) of {\it individual}\/ objects --- individual particles,
black holes, \etc\ --- within the system, or on the dynamical variables 
$\Omega_i$ and $H$, or on $t$ itself.  In other words, the pump rate 
was independent of the external environment in which these constituent objects 
found themselves.

In this paper, by contrast, we have demonstrated that a pump with the properties appropriate 
for stasis can also arise from a wholly different class of 
physical process --- processes in which this rate depends on additional dynamical 
variables which characterize the {\it extrinsic}\/ properties of a cosmological population 
of particles, black holes, or other individual objects.  In other words,
in such cases, the pump rates depend not only on the intrinsic properties of these 
constituent objects, but also on their external cosmological environment.
We have shown, for example, that the annihilation of a gas consisting of a single 
species of non-relativistic matter particle into radiation --- a 
process whose rate depends not only on the overall abundance $\Omega_M$ of these matter 
particles, but also on their temperature $T$ --- can, under the certain conditions, give rise 
to such a pump.  Notably, unlike in all realizations of stasis which 
have previously been identified in the literature, stasis emerges in this context 
in a manner which does not require a tower of states.
Moreover, we have shown that stasis nevertheless emerges as a dynamical attractor 
in systems wherein these conditions are satisfied.

The principal such condition is that the swept-volume rate $\sigma v$ for the annihilation
process must scale in an appropriate manner with the momenta of the annihilating 
matter particles in the CM frame.  While this condition is a non-trivial one, we 
have detailed a particle-physics mechanism through which such a momentum-dependence
can be achieved.

Several additional comments are in order.  First, 
as we have already noted, thermal stasis --- unlike its non-thermal cousins --- involves 
thermal effects in an intrinsic way, \ie, as part of the pump that establishes and sustains 
the resulting stasis.   That said, we find it remarkable that this stasis does not simply 
keep the temperature of our $\phi$-particle gas constant, as might naively been anticipated.  
Such a result, of course, would have rendered a thermal realization of stasis challenging to 
realize in an expanding universe, since additional dynamics would be required in order to 
maintain this temperature at a constant value in such a context. 
However, what we find is that our thermal stasis also has a continually dropping temperature!
That said, the rate at which the temperature drops is modified during thermal stasis, 
due to the manner in which annihilation affects different parts of the phase-space distribution 
of the $\phi$ particles.  In particular, the universe cools slightly more slowly than one 
would expect as a result of expansion alone.
Indeed, what we have a discovered is that there is an entirely {\it new}\/ thermodynamic 
quantity, the {\it coldness}\/ $\Xi$, which remains fixed during stasis.
This observation also suggests that it is coldness, rather than temperature, which may be 
a more fundamental dynamical variable as far as stasis is concerned.

Second, despite the absence of a tower of states, the realization of thermal 
stasis that we have described in Sect.~\ref{sec:mechanism_q_neg_2} comes with 
its own graceful exit.   In tower-based realizations of stasis, the stasis epoch 
stretches across a time interval during which the action of the pump incrementally 
proceeds towards the lower portions of the 
tower~\cite{Dienes:2021woi,Dienes:2023ziv,Dienes:2024wnu}.  The exit from the stasis 
epoch then arises once we reach the bottom of the tower.  As a result, once stasis ends,
the universe subsequently becomes dominated by that cosmological component
which participates in the stasis dynamics and  which has the highest equation-of-state 
parameter.  For thermal stasis, by contrast, 
there is no tower.  We nevertheless continue to have an extended stasis epoch, 
and this in turn ends through a different mechanism:
once the temperature of our $\phi$-particle gas
eventually drops below the temperature $\Tmin$ in Eq.~(\ref{eq:T_bounds_propagator}), 
$\langle \sigma v\rangle$ no longer increases as the kinetic-energy density 
of the $\phi$ particles decreases. 
As a result, $\phi\phi\to\chi\chi$ annihilation becomes inefficient and $\Omega_M$ 
increases until the $\phi$-particle gas (\ie, the component with the {\it lowest}\/ 
equation-of-state parameter) ultimately comes to dominate the energy density of 
the universe.  Thus, in this scenario, stasis ends not with a bang, but with a WIMP-er.
Clearly the graceful exits for thermal and tower-based 
realizations of stasis are qualitatively quite different.  We nevertheless see that 
each of our stasis epochs comes with its own intrinsic exit.

%==================
\begin{figure}[t]
\centering
\includegraphics[keepaspectratio, width=0.49\textwidth]{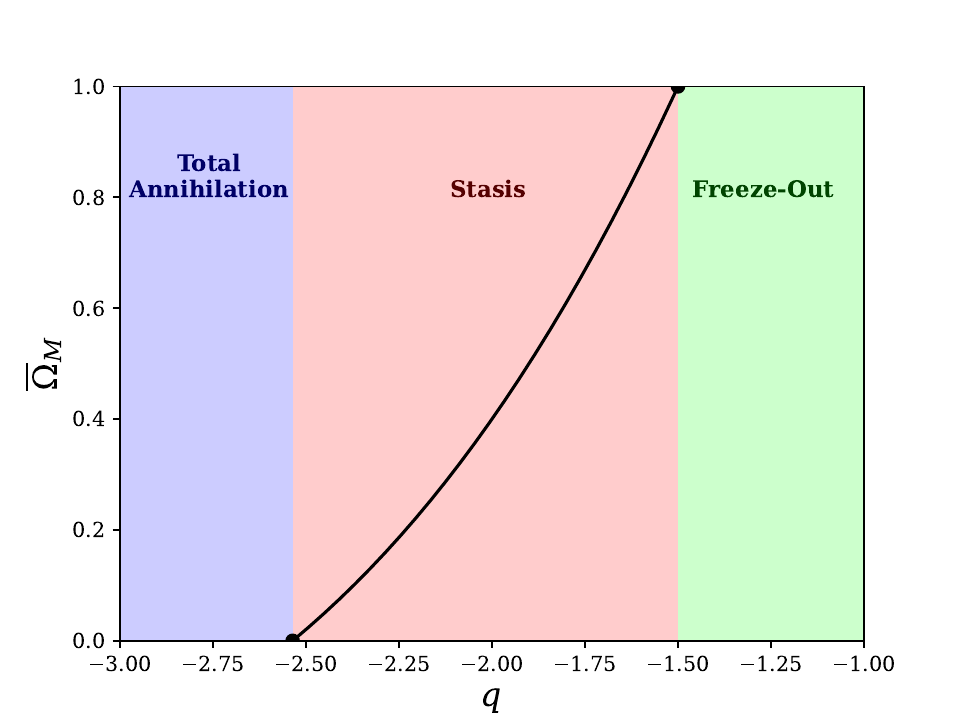}
  \caption{The stasis abundance $\barOmega_M$, plotted as a function of $q$ within 
  the range $-6 + 2\sqrt{3} < q < -3/2$ (pink background) for which thermal stasis 
  can be achieved.  For $q < -6+2\sqrt{3}$ (blue background), the swept-volume rate 
  $\sigma v$ remains sufficiently high for arbitrarily small $|\vec{p}_{\rm CM}|$ --- and 
  thus at arbitrarily large distance scales --- that the population of $\phi$ particles 
  annihilates completely.  By contrast, for $q > -3/2$ (green region), this swept-volume 
  rate becomes sufficiently small for small $|\vec{p}_{\rm CM}|$ that $\phi$ particles 
  cannot find each other at late times and therefore effectively freeze out.  Thermal 
  stasis may thus be interpreted as a phenomenon which interpolates between these two 
  regimes, with a swept-volume rate that decreases sufficiently rapidly with increasing 
  $|\vec{p}_{\rm CM}|$ that total annihilation is avoided, and yet  not so rapidly that a 
  relic population of $\phi$ particles freezes out at low temperatures.   
  \label{fig:Wbar_of_q2}
  }
\end{figure}
%==================

On a final note, we observe that 
the manner in which $\barOmega_M$ varies with $q$ across the range
specified in Eq.~(\ref{eq:q-range}) suggests that this stasis window can be 
viewed as interpolating between a regime wherein our population of $\phi$ 
particles annihilates completely and a regime wherein this population of 
particles effectively freezes out.  
It is straightforward to understand why this is the case.
In Fig.~\ref{fig:Wbar_of_q2}, we have schematically indicated a range of $q$-values that 
not only includes the window $-6+2\sqrt{3}<q< -3/2$ in which stasis emerges (shaded in pink), 
but also extends beyond this window in each direction (blue and green). 
We can then discuss what happens within each region in turn.

For values of $q$ below this stasis window (\ie, within the blue-shaded region), the 
swept-volume rate $\sigma v$ remains sufficiently high for arbitrarily 
small $|\vec{p}_{\rm CM}|$ --- and thus at arbitrarily large distance 
scales --- that the population of $\phi$ particles annihilates completely.  
As a result, there is no stasis in the blue-shaded region:  the pump is overwhelmingly 
effective for such values of $q$, and $\Omega_M$ simply falls to zero.  

At the opposite extreme, for values of $q$ above the stasis window (\ie, within the 
green-shaded region), $\sigma v$ becomes 
sufficiently small for small $|\vec{p}_{\rm CM}|$ that $\phi$ particles cannot 
find each other at late times.  A situation then develops which is in many
ways similar to the situation that arises when a WIMP freezes out.  
While there are of course salient differences 
between the freeze-out mechanics of a WIMP and the freeze-out mechanics of our 
$\phi$-particle gas --- for example, the $\phi$ particles are never in thermal 
equilibrium with the radiation particles into which they annihilate --- the 
qualitative result is the same:  a relic population of $\phi$ 
particles is left over once these particles effectively cease annihilating with
each other, and this population of $\phi$ particles comes to dominate the energy 
density of the universe at late times, with $\Omega_M$ becoming effectively unity.  
Thus once again no stasis is possible.
Indeed, we note that this phenomenon is precisely how the graceful exit from stasis 
discussed earlier comes to pass: below $\Tmin$, the value of $q$ is effectively 
no longer $q=-2$ but instead $q=0$, and thus our population of matter particles 
freezes out. 

Finally, between these two extremes lies our stasis window (shaded in pink).  
For $q$ within this window, $\sigma v$ decreases with increasing $|\vec{p}_{\rm CM}|$ 
sufficiently rapidly that total annihilation is avoided, but not so 
rapidly that a relic population of $\phi$ particles freezes out at low temperatures. 
Thus a {\it bona fide} stasis emerges, with corresponding  values of $\barOmega_M$ that 
interpolate between the two extremes, as illustrated in Fig.~\ref{fig:Wbar_of_q2}.  
It is noteworthy --- and even somewhat remarkable --- that this transition between 
the two extremes outlined above does not occur abruptly at some critical value of $q$, 
but rather over a non-zero range of $q$-values, with stasis emerging for all values of 
$q$ within this window and with the corresponding stasis abundance of the 
$\phi$-particle gas varying smoothly over the entire allowed range 
$0\leq\barOmega_M \leq 1$ as we traverse this window.  Indeed, we see that the balancing 
inherent in stasis can be achieved for {\it any}\/ $q$-value within the stasis window, 
with the resulting stasis abundance $\barOmega_M$ varying with the specific value of $q$. 

Given our results, many avenues are now open for future research.  For example, in
Sect.~\ref{sec:mechanism_q_neg_2} we have developed a particle-physics mechanism for 
achieving a swept-volume rate of the form in Eq.~(\ref{eq:cross1}) with $q=-2$.  However, it
nevertheless remains to construct a concrete particle-physics {\it model}\/ in which this 
mechanism is realized in a self-consistent way.
Of course, any concrete model along these lines must not only yield the desired 
momentum-dependence for $\sigma v$ across a significant range of $|\vec{p}_{\rm CM}|$-values, 
but must also satisfy a number of additional self-consistency conditions and 
observational constraints.  First steps toward the construction of such a model can be 
found in Ref.~\cite{Barber_toappear_model}.~  It will also be interesting to determine 
the specific observational signatures to which a thermal stasis epoch might lead.    
Work in all of these directions is underway.

%%%%%%%%%%%%%%%%%%%%%%%%%%%%%%%%%%%%%%%%%%%%%%%%%%%%%%%%%%%%%%%%%%%%%%%%%%%%%%
\bigskip
\bigskip

\begin{acknowledgments}

We are happy to thank L.~Heurtier, F.~Huang, T.~M.~P.~Tait, and U.~van~Kolck for 
discussions.  The research activities of 
JB and KRD are supported in part by the U.S.\ Department of Energy under Grant 
DE-FG02-13ER41976 / DE-SC0009913; the research activities of KRD are also 
supported in part by the U.S.\ National Science Foundation through its employee 
IR/D program.  The research activities of BT are supported in part by the 
U.S.\ National Science Foundation under Grants PHY-2014104 and PHY-2310622.  
The opinions and conclusions expressed herein are those of the authors, and do 
not represent any funding agencies. 

\end{acknowledgments}

\bibliography{TheLiterature2}

\end{document}